\documentclass[aip,reprint]{revtex4-1}

\usepackage{multirow}
\usepackage{graphicx}
\usepackage{xspace}

\newcommand{\ie}{i.e.,\xspace}
\newcommand{\eg}{e.g.,\xspace}
\newcommand{\lc}{\left [}
\newcommand{\rc}{\right ]}
\newcommand{\lp}{\left (}
\newcommand{\rp}{\right )}

\begin{document}


{\footnotesize This article may be downloaded for personal use only. Any other use requires prior permission of the author and AIP Publishing. This article appeared in L. A. Martinez-Vaquero, "Inequality leads to the evolution of intolerance in reputation-based populations", Chaos \textbf{33} (3), 033119 (2023) (doi.org/10.1063/5.0135376) and may be found at https://pubs.aip.org/aip/cha/article-abstract/33/3/033119/2881304/Inequality-leads-to-the-evolution-of-intolerance.
	
Copyright 2014 Luis A. Martínez-Vaquero. This article is distributed under a Creative Commons Attribution License.}
	
\title{Inequality leads to the evolution of intolerance in reputation-based populations} 



\author{Luis A. Martinez-Vaquero}
\email[]{l.martinez.vaquero@upm.es}
\affiliation{Grupo de Sistemas Complejos, Universidad Politécnica de Madrid, 28040 Madrid, Spain}
\affiliation{DEFE, Escuela Técnica Superior de Arquitectura, Universidad Politécnica de Madrid, 28040 Madrid, Spain}
\affiliation{Grupo Interdisciplinar de Sistemas Complejos, Universidad Carlos III de Madrid, 28911 Legan\'es, Madrid, Spain}



\begin{abstract}
 This work studies the impact of economic inequality on the evolution of intolerance through a reputation-based model of indirect reciprocity. Results show that economic inequality is a powerful enhancer of intolerance, inducing the escalation of out-group discrimination even without the presence of new intolerant mutants. It also generates behavior modifications within tolerant disfavored minorities: their members either relax punishments against the uncooperative or prioritize helping the wealthy, even suffering discrimination in return. On the other hand, the redistribution of wealth is proved as a viable solution to avoid the spread of intolerance as long as it increases equality and is implemented before intolerance permeates part of the population.
\end{abstract}

\maketitle 

\begin{quotation}
Economic inequality has been rising in modern societies during the last years and, according to evidence, it could be a key factor behind intolerance outbreaks. As our societies become more dependent on public reputation systems, a deeper understanding of the relationship between inequality, intolerance, and reputation is required. Evolutionary game theory has been proved as a suitable methodology to study the emergence and evolution of cooperation in populations. One of the main mechanisms proposed to explain cooperative behaviors in this framework is indirect reciprocity, which relies on the assignment of reputations to decide whether to cooperate with the other. Adapting an indirect reciprocity model to incorporate economic inequality and intolerance, this work studies the impact of the former on the emergence of the latter. It is analyzed under what circumstances intolerant strategies could invade tolerant groups in different scenarios and what is required for tolerance to be restored. Results show that inequality triggers the escalation of discriminatory attitudes at different levels. However, it is also responsible of counterintuitive behaviors within tolerant groups. One of the most significant is that tolerant disfavoured minorities may prioritize helping  wealthy individuals who discriminate against over their own kind. The model is also applied to study the conditions under which the redistribution of wealth may become a feasible solution to avoid intolerance. Results show that this policy should guarantee the reduction of inequality and be implemented before intolerant individuals invade any group of the population.

\end{quotation}


\section{Introduction}

\begin{table*} 
		\caption{\label{tab:table1}Summary of the main magnitudes used in the model. }
		\begin{ruledtabular}
			\begin{tabular}{ll}
				Magnitude & Description\\
				\hline
				$b$ & Benefit of receiving help. \\
				$c$ & Cost of helping. \\
				$a_{\alpha,\beta}$ & Action rule: decision of a donor with reputation $\alpha$ on helping or not a recipient with reputation $\beta$. \\
				$m_{\alpha,\beta}(a)$ & Moral assessment: new reputation assigned to an individual with reputation $\alpha$ which has performed\\
				& an action $a$ on an individual with reputation $\beta$.\\
				$y$ & Fraction of individuals belonging to subpopulation A.\\
				$\epsilon_X$ & Economic stress of subpopulation $X$.\\
				$\epsilon$ & Global economic stress.\\
				$G$ & Gini coefficient. \\
				$\eta_{X,Y}$ & Intention of cooperation of individuals from subpopulation $X$ over individuals from subpopulation $Y$. \\
				$\theta_{X,Y}$& Probability that a $X$-strategist helps a $Y$-strategist. \\
				$W_i$ & Average payoff that $i$-individuals receive. \\ 
				$x_i^{\Lambda_A \Lambda_B \Lambda_M}$ & Fraction of $i$-individuals with reputations
				$\Lambda_A$, $\Lambda_B$, and $\Lambda_M$ assigned by A-residents, B-residents, and
				mutants. 
			\end{tabular}
			\label{tab_sum}
		\end{ruledtabular}
	\end{table*}

The evolution of inequality from small-scale egalitarian societies was a major
transition in human social organization. It emerged in Holocene, when resources became predictable and followed gradient or patching distributions \cite{mattison2016evolution}.
During the last decades, economic inequality has been decreasing among nations through the improvement of low-income countries' economies. However, the picture within nations is pretty different: the  gaps between higher and lower income economic classes have been continuously growing \cite{Alvaredo+2018}.  Some authors argued that globalization has been one of the main catalysers of inequality \cite{mazur2000labor}.
Modern democracies have seen how their income inequality has skyrocketed in recent decades \cite{firebaugh2000trend, goesling2001changing,  fischer2006century, piketty2018distributional}. Even European countries, which have developed policies orientated to ensure a fairer share of the economic growth among the low level income groups, have failed to achieve the inequality reduction United Nations Sustainable Development Goal \cite{blanchet2019unequal}.
Moreover, national average income levels have shown not to be good proxies of inequality for most countries since some high- or middle-income countries display extreme inequality levels, such as Brazil, India and China \cite{chancel2021world}.

On the other hand, discriminatory behaviors are relatively common in heterogeneous societies where 
individuals
tend to form differentiated groups
\cite{sumner:1906, levine:1972, Hirschfeld:1996, bernhard:2006}.
Discrimination usually comes in two different flavours \cite{struch:1989}: intolerance against individuals from other groups \cite{allport1954nature, fiske2000stereotyping, fiske2002we} and favoritism toward same group individuals \cite{yamagishi1999bounded, fu2012evolution, balliet2014ingroup}. Intolerance can be materialized through many different discriminatory behaviors, mainly related to ethnicity, gender, age, sexual preferences, religion and nationality \cite{ray:1986, palmore1999ageism, cashdan:2001, bertrand2017field, ellemers2018gender}.
Despite the fact that the relationship between economic income  has been extensively discussed and analyzed 
\cite{dollard:1939, hovland:1940, lipset1959democracy, levin:1982, persell2001civil,  gibson:2002, svallfors2006moral}, 
some evidences indicate
that economic inequality may be at least as important as the absolute economic income in social trust and tolerance \cite{knack1997does, uslaner2002moral, andersen2008economic, andersen2012polarizing, nhim2019resilience}.

Previous theoretical works have shown that discriminatory behaviors can evolutionarily emerge in the framework of
indirect reciprocity \cite{masuda:2012, nakamura:2012, oishi:2013, whitaker2018indirect} and, more specifically, they can outbreak more easily under global economic stress especially among minorities in reputation-based populations \cite{martinez:2014}.
Indirect reciprocity  \cite{sugden:1986, alexander:1987} is one of the main mechanisms proposed to explain the evolution of cooperation under the evolutionary game theory framework \cite{nowak:1998, boyd1989evolution, leimar:2001, panchanathan:2003, nowak:2005, nowak:2006b}, demonstrating its relevance in
human societies \cite{dufwenberg:2001, milinski:2002, panchanathan:2004,
	semmann:2004, suzuki:2007, yoeli2013powering}.
Contrary to direct reciprocity \cite{trivers:1971}, where individuals react to their partners' previous actions, indirect reciprocity involves a third party. There exist different variants of indirect reciprocity \cite{boyd1989evolution, nowak:2007}, however the most prevalent is based on reputation  \cite{nowak:1998, ohtsuki:2004, brandt:2004, fehr2004don, nowak:2005,
	milinski:2002}.
Individuals judge others by assigning them reputations based on the actions they witness, which in turn influence decisions in future encounters.
The emergence of moral assessments is an interesting consequence of
these models which have received significant attention in the literature \cite{ohtsuki:2004,ohtsuki:2006b,martinez:2013}, making them suitable to study the conditions under which intolerant
behaviors can appear and spread in a population.
Traditionally, the importance of reputation mechanisms was restricted mainly to small-scale societies, however, as a result of the proliferation of public online reputation-based systems, it influences decisions at all scale levels in current societies \cite{bolton2004effective, josang2007survey}.

Despite all this progress, the role of inequality still remains unclear in the evolutionary emergence of intolerance in general and in reputation-based societies in particular.
The present work proposes an evolutionary game theoretical model based on indirect reciprocity where inequality is incorporated considering different groups with different economic levels. 
This allows to
study how inequality impacts the intention of cooperation within and across groups and how it shapes the evolution of intolerant behaviors.
It is also analyzed under what conditions the redistribution of wealth may become a potential solution to contain the emergence of intolerance.

\section{Model}

The model proposed in this work is built upon previous reputation-based
models of indirect reciprocity \cite{brandt:2004, martinez:2014} to
account for the effect of group economic inequality on the emergence of intolerance.
Table~\ref{tab_sum} summarizes the main magnitudes used in this model.
It is assumed an infinitely
large, well-mixed population of individuals who randomly interact in pairs, involving a \emph{donor} and a \emph{recipient.} The
donor can either help the recipient (C from \emph{cooperate}), giving an amount $b$ at a personal cost $c$
($b>c$), or fail to do it (D
from \emph{defect}).
The whole population witness the action performed, and each individual
judges this action either as \emph{good} (G) or as \emph{bad} (B), updating the donor's reputation if necessary. Therefore everyone in
the population
has an opinion regarding every other one’s attitudes.
Assuming third-order strategies, the donor's decision to
help or not to help the recipient depends on her reputation $\alpha$, and the reputation
of the recipient $\beta$ (both judged from the viewpoint of the donor) according to \textit{action rules} $a_{\alpha\beta}$.
Furthermore, every individual judges any action witnessed $a$ according to
\textit{moral assessments} $m_{\alpha\beta}(a)$, which involves the reputations of the donor
and the recipient (from the viewpoint of the witness) and the action performed.
Pairwise interactions between members of the population continue to happen
indefinitely. Thus, individuals' payoffs are computed as the average payoffs of
all interactions they are involved in once the equilibrium of the dynamics is
reached.

The emergence of intolerance is allowed by dividing the
population into two different groups: a group A with $y$ fraction of individuals and a group B with $1-y$ individuals.
According to their behavior toward the other group, individuals can be
classified as tolerant or intolerant.
Tolerant individuals always stick to the chosen strategy ignoring individuals' groups
---they help or not, or judge, as if everyone
belonged to the same group. Intolerant individuals, on the contrary, always
assign bad reputation to individuals from the opposite group and 
refuse to help them;
they also assign
bad (good) reputation to anyone who helps (refuses to help) someone
from the opposite group. Other than that, all individuals in the population
share the same action rules and moral assessments.

The model introduces economic stress by limiting individuals' resources 
imposing a fraction of failed assistances, which is referred to in the literature as
\textit{action error} \cite{leimar:2001, ohtsuki:2004, panchanathan:2003,
	fishman:2003, lotem:1999}.
In order to analyze how
economic inequality affects the spread of intolerance, A and
B individuals are allowed to undergo different levels of economic stress
$\epsilon_A$ and $\epsilon_B$, respectively. 
The standard index to measure economic inequality is the Gini coefficient \cite{gini1936measure}. For
a population with two levels of income,
it is
simply defined as the difference between the fraction of income of the
wealthiest subpopulation, minus the fraction of the total population that
correspond to the wealthiest people. Taking as reference subpopulation A, and $\epsilon_X$ being the economic stress of subpopulation $X$, it is
reasonable to assume that subpopulation A's income is proportional to
$y(1-\epsilon_{{A}})$, and that the global income is
$1-\epsilon=y(1-\epsilon_{{A}})+(1-y)(1-\epsilon_{{B}})$. Hence the Gini
coefficient becomes
\begin{equation}
	G= \frac{y(1-y)}{1-\epsilon}(\epsilon_{{B}}-\epsilon_{{A}}) 
	\label{eq:Gini}
\end{equation}
The value of the economic stress of each subpopulation can be recovered from the Gini coefficient for a given global economic stress $\epsilon$:

\begin{eqnarray}
		\epsilon_{A} = \epsilon - \frac{1-\epsilon}{y}\, G \\
		\epsilon_{B} = \epsilon + \frac{1-\epsilon}{1-y}\, G 
\end{eqnarray}
Thanks to this artefact it is easy to interpret the meaning of increasing or
decreasing $G$ either way. The higher the absolute value of $G$, the more unequal the population in favor of subpopulation A ($G>0$) or subpopulation B ($G<0$).

\begin{figure}[!t]
	\includegraphics{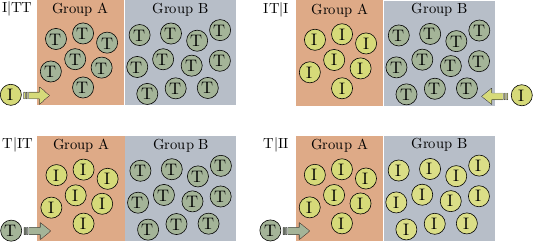} 
	\caption{Conceptual representation of the four possible invasions analyzed: intolerance invading a tolerant group in fully tolerant population (I\textbar TT), tolerance invading an intolerant group with other group remaining tolerant (T\textbar IT), intolerance invading a tolerant group in a population where the other group has already become intolerant (IT\textbar I), and tolerance invading a intolerant group in a fully intolerant population (T\textbar II). }
	\label{fig:invas}
\end{figure}

\begin{table*}
	\caption{Leading eight strategies. Moral assessments $m_{\alpha\beta}(a)$ and action rules $a_{\alpha\beta}$ for third-order indirect reciprocity strategies that are evolutionarily stable and obtain high payoff when in homogeneous populations. Strategies are arranged in three different groups according with their similar behavior. Strategies in bold are used as representatives of each group for subsequent analysis.}
	\begin{ruledtabular}
		\begin{tabular}{ccccccccccccc}
			& $m_{GG}(C)$ & $m_{GG}(D)$ & $m_{GB}(C)$ & $m_{GB}(D)$ & $m_{BG}(C)$ &
			$m_{BG}(D)$ & $m_{BB}(C)$ & $m_{BB}(D)$ & $a_{GG}$ & $a_{GB}$ & $a_{BG}$ & $a_{BB}$ \\
			\hline
			\multirow{2}{*}{\textbf{I}}  & \textbf{G} & \textbf{B} & \textbf{B} & \textbf{G} & \textbf{G} & \textbf{B} & \textbf{G} & \textbf{B} & \textbf{C} & \textbf{D} & \textbf{C} & \textbf{C}   \\
			& G & B & G & G & G & B & G & B & C & D & C & C  \\
			
			\hline
			\multirow{4}{*}{\textbf{II}} & \textbf{G} & \textbf{B} & \textbf{G} & \textbf{G} & \textbf{G} & \textbf{B} & \textbf{B} & \textbf{G} & \textbf{C} & \textbf{D} & \textbf{C} & \textbf{D}  \\
			& G & B & G & G & G & B & G & G & C & D & C & D \\
			& G & B & B & G & G & B & G & G & C & D & C & D  \\
			& G & B & B & G & G & B & B & G & C & D & C & D  \\
			\hline
			\multirow{2}{*}{III}& G & B & G & G & G & B & B & B & C & D & C & D  \\
			& G & B & B & G & G & B & B & B & C & D & C & D  \\
		\end{tabular}
		\label{tab_leading8}
	\end{ruledtabular}
\end{table*}

Among all possible strategies ---combinations of action rules and moral assessments---, the so-called \textit{leading eight strategies} (see
Table~\ref{tab_leading8}) were proved evolutionarily stable,
obtain significantly high payoffs,
promote cooperation and are robust against errors and cheating \cite{ohtsuki:2004, martinez:2013}. 
For each of the leading eight strategies, 
it was
analyzed the range of parameters for which different invasions may succeed (see next sections for details). 
The different invasion scenarios are represented in Fig.~\ref{fig:invas} 
and  denoted $S_M|S_A S_B$ or $S_A S_B|S_M$, depending on whether it is the A or the B subpopulation that is being invaded by mutants M, respectively, where $S_i$ can take the values T (tolerant strategy) or I (intolerant strategy).
Thus, e.g., I\textbar TT means that intolerant mutants are attempting to
invade the A subpopulation of a fully tolerant population; or TI\textbar T
means that tolerant mutants are attempting to invade the B subpopulation of a
population formed by A tolerant individuals and B intolerant individuals.
The leading eight strategies are usually classified in three groups (see
Table~\ref{tab_leading8}) according to their stability against invasions by
free-riders
\cite{ohtsuki:2004}. Since it has already been shown that strategies of the same group behave similarly and strategies of group \textbf{III} resist invasions by intolerant strategies under any
circumstance \cite{martinez:2014},
most results displayed from now on will refer to one representative strategy of each group \textbf{I} and \textbf{II}.

\subsection{Evolutionary stability}

Each invasion scenario analyzed consists of a resident population and a small fraction of mutant individuals that try to invade one of the subpopulations. Resident populations are prepared by making all combinations of A tolerant/intolerant individuals with B tolerant/intolerant ones; mutants exhibit the opposite (in)tolerant behavior of the subpopulation they are attempting to invade.
Only the onset of the
invasion is determined, but not the final composition of the population. Finding whether the
invading strategy replaces the resident population or end up coexisting with it
is a much more complex problem \cite{martinez:2014} and is beyond the scope of this work.

A strategy is evolutionarily stable if a population fully formed by individuals who play that strategy can resist the invasion of a small fraction of mutants that play other strategies \cite{smith1973logic}. Assuming a population formed by A-type and B-type residents, the condition for the former to resist the invasion of mutants that play a different strategy is
$W_A>W_M$, where $W_A$ and $W_M$ are the average payoffs received
by A-type residents and by mutants, respectively. They can be obtained as
\begin{eqnarray}
		W_A&&=b \left[y\ \theta_{A,A}+(1-y)\ \theta_{B,A} \right]\ \nonumber\\
		&&-c \left[y\ \theta_{A,A}+(1-y)\ \theta_{A,B} \right], 
	\label{eq:WA1_gen}
\end{eqnarray}
\begin{eqnarray}
		W_M&&=b \left[y\  \theta_{A,M}+(1-y)\  \theta_{B,M} \right] \nonumber\\
		&&-c \left[y\  \theta_{M,A}+(1-y)\ \theta_{M,B} \right], 
	\label{eq:WM1_gen}
\end{eqnarray}
where $\theta_{X,Y}=(1-\epsilon_X)\eta_{X,Y}$ is the probability that a $X$-strategist helps a
$Y$-strategist. The intention of cooperation $\eta_{X,Y}=0$ if the $X$-strategist is intolerant and $X\neq Y$, and
\begin{equation}
	\eta_{X,Y}=\sum_{\alpha \beta} \chi_\alpha(g_X^X) \chi_\beta(g_Y^X) a_{\alpha \beta}
\end{equation}
otherwise, where $g_i^j$ is the fraction of $i$-strategists that are considered good by
$j$-strategists and $\chi_{\gamma}(x)=\gamma x+(1-\gamma)(1-x)$.
In the case where mutants attempt to invade the B subpopulation, B-type residents resist the invasion if $W_B>W_M$ interchanging subscript A by B and $y$ by $1-y$  in Eqs.~\ref{eq:WA1_gen} and~\ref{eq:WM1_gen}:
\begin{eqnarray}
		W_B=&&b \left[(1-y)\ \theta_{B,B}+y\ \theta_{A,B} \right]\ \nonumber\\
		&&-c \left[(1-y)\ \theta_{B,B}+y\ \theta_{B,A} \right],
\end{eqnarray}
\begin{eqnarray}		
		W_M=&&b \left[(1-y)\  \theta_{B,M}+y\  \theta_{A,M} \right] \nonumber\\
		 &&-c \left[(1-y)\  \theta_{M,B}+y\ \theta_{M,A} \right], 
	\label{eq:WB1_gen}
\end{eqnarray}

\subsection{Reputation dynamics}

Defining $x_i^{\Lambda_A \Lambda_B \Lambda_M}$ as the fractions of $i$-individuals with reputations
$\Lambda_A$, $\Lambda_B$, and $\Lambda_M$ assigned by A-residents, B-residents, and
mutants, respectively, fractions $g_i^j$ can be expressed as
\begin{equation}
	g_{i}^{A} = x_i^{G**}, \qquad
	g_{i}^{B} = x_i^{*G*}, \qquad
	g_{i}^{M} = x_i^{**G},
\end{equation}
where the superscript * stands for \textit{any reputation}. For instance, $x_i^{G**} =  \sum_{\Lambda_B} \sum_{\Lambda_M} x_i^{G \Lambda_B \Lambda_M}$.
In practice, only 
a few of these fractions
are necessary to compute the different $g_i^j$.
Since mutants appear in a very low rate, donors only interact with residents. Under this approximation, the dynamics of the fractions of residents are reduced to 
\begin{eqnarray}
	 \dot{x}_{A}^{\Lambda_A*\Lambda_M} &=y\ T_{A,A}^{\Lambda_A\Lambda_M} +
	(1-y)\ F_{A,B}^{\Lambda_A\Lambda_M} - x_A^{\Lambda_A*\Lambda_M}, \label{eq:evol_xA_gen}\ \ \  \\
	\dot{x}_{A}^{\Lambda_A\Lambda_B*}&=y\ T_{A,A}^{\Lambda_A\Lambda_B} +
	(1-y)\ F_{A,B}^{\Lambda_A\Lambda_B} - x_A^{\Lambda_A\Lambda_B*},  \label{eq:evol_xA1_gen}\ \ \  \\
	\dot{x}_{B}^{\Lambda_A\Lambda_B*}&= y\ F_{B,A}^{\Lambda_A\Lambda_B} +
	(1-y)\ T_{B,B}^{\Lambda_A\Lambda_B}  - x_B^{\Lambda_A\Lambda_B*},  \label{eq:evol_xB1_gen}\ \ \  \\
	\dot{x}_{B}^{*\Lambda_B\Lambda_M}&= y\ F_{B,A}^{\Lambda_B\Lambda_M} +
	(1-y)\ T_{B,B}^{\Lambda_B\Lambda_M} - x_B^{*\Lambda_B\Lambda_M}, \label{eq:evol_xB_gen}\ \ \ 
\end{eqnarray}
and are decoupled from the equations that governs the dynamics of the mutants, which are given by:
\begin{eqnarray}	
	\dot{x}_{M}^{\Lambda_A*\Lambda_M}&=y\ T_{M,A}^{\Lambda_A\Lambda_M} +
	(1-y)\ F_{M,B}^{\Lambda_A\Lambda_M} - x_M^{\Lambda_A*\Lambda_M}, \label{eq:evol_xM_gen}\ \ \ \\
	\dot{x}_{M}^{*\Lambda_B\Lambda_M}&= y\ F_{M,A}^{\Lambda_B\Lambda_M} +
	(1-y)\ T_{M,B}^{\Lambda_B\Lambda_M} - x_M^{*\Lambda_B\Lambda_M} \label{eq:evol_xMB_gen}\ \ \
\end{eqnarray}
in the case of A-type and B-type mutants, respectively. In previous Eqs.~\ref{eq:evol_xA_gen}--\ref{eq:evol_xMB_gen}, the interactions with recipients of the same and of the opposite group
have been split into $T_{i,j}^{\Lambda_l\Lambda_m}$ and $F_{i,j}^{\Lambda_l\Lambda_m}$,
respectively. Fractions $T_{i,j}^{\Lambda_l\Lambda_m}$ are common for all invasion scenarios whereas  $F_{i,j}^{\Lambda_l\Lambda_m}$ depend on the specific scenario. These fractions together with other further simplifications and the processes to complete calculations are detailed in Appendix.

\section{Results}

\subsection{Intention of cooperation}

\begin{figure*}[!t]
	\includegraphics[width=7in]{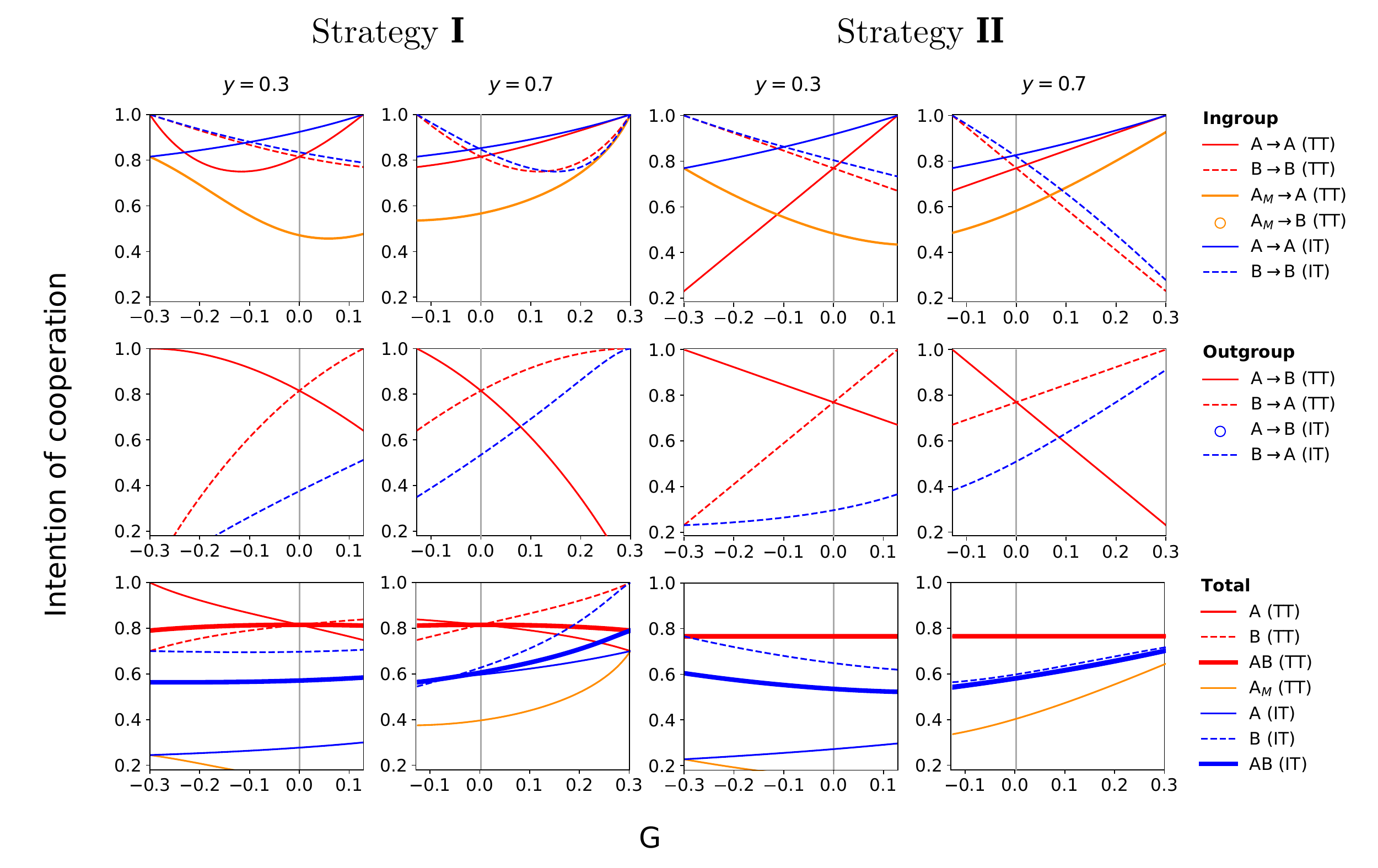} 
	\caption{Intention of cooperation under inequality. Intention of cooperation level of subpopulations X with subpopulations Y ($\text{X}\rightarrow \text{Y}$) as a function of the Gini coefficient for a completely tolerant population (TT) and a population where A-individuals became intolerant (IT). Intention of cooperation levels are grouped according to in-group (first row) and out-group (second row) behavior criteria. Total intention of cooperation for each subpopulation and the population as a whole is also shown (third row). Results for different subpopulation sizes and strategies are displayed. Empty circles in the legend stand for cases where the intention of cooperation is always zero, and vertical grey lines mark equality ($G=0$). Global economic stress $\epsilon=0.3$. }
	\label{fig:coop}
\end{figure*}

\begin{figure*}[!t]
	\includegraphics[width=5in]{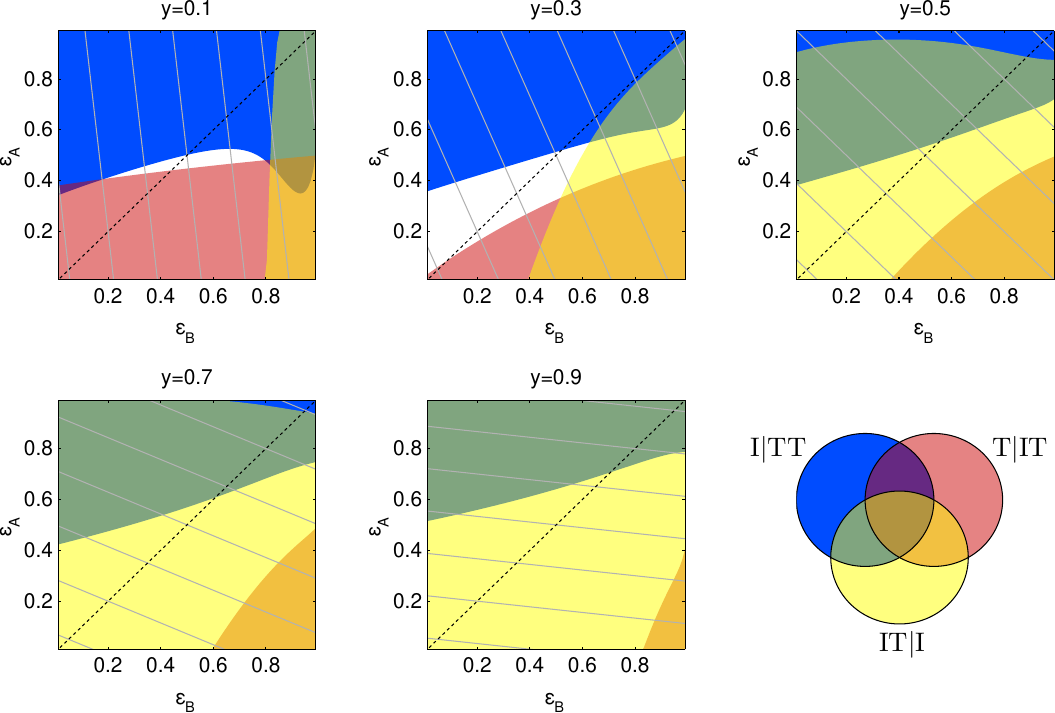} 
	\caption{{Invasion of intolerance for strategy I}. Different regions of the $\epsilon_A$--$\epsilon_B$ plane where invasions I\textbar TT 
		(blue), T\textbar IT (red) and T\textbar IT (yellow) are successful for strategy \textbf{I} and
		different fractions of subpopulation A. Overlaps are colored with the
		corresponding mixed color (the key is provided by the overlapping circles). Grey
		continuos lines correspond to constant global economic stress
		$\epsilon=y\,\epsilon_A+(1-y)\epsilon_B$. Gini coefficient changes moving along these lines from $G=0$ on the diagonal $\epsilon_A=\epsilon_B$ (dotted lines) to $G>0$ below
		it (favoring subpopulation A) or to $G<0$ above it (favoring subpopulation B).
		Benefit-to-cost ratio $b/c=2$.}
	\label{fig:I}
\end{figure*}

\begin{figure*}
	\includegraphics[width=5in]{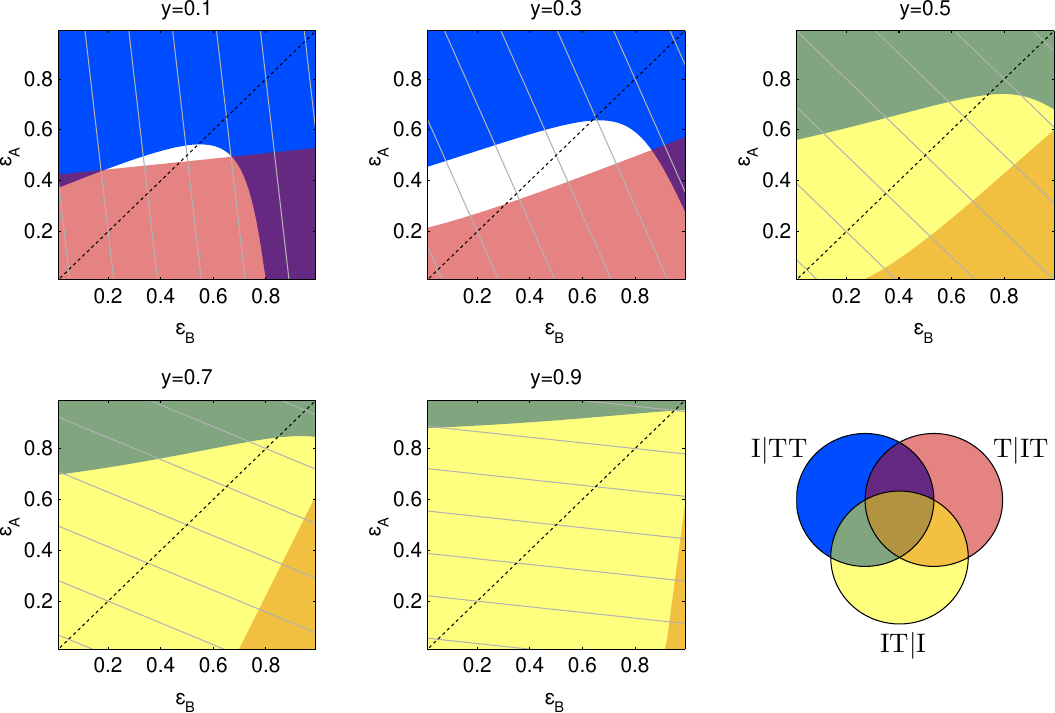}
	\caption{{Invasion of intolerance for strategy II}. Notation and color criteria follow the same as in Fig.~\ref{fig:I}.}
	\label{fig:II}
\end{figure*}

The study of the cooperation level as a function of inequality under different configurations provides first insights on why intolerance can or cannot outbreak in fragmented societies. Specifically,
it will be focused
on the intention of cooperation, which is the probability that an individual from subpopulation X \textit{intends} to cooperate with an individual from subpopulation Y according only to her strategy.
This magnitude is more adequate to infer the actual individuals' intentions regardless of their resource limitations. Note that economic stress still has an important impact on the intention of cooperation since it influences individuals' reputations, which are based on others' actual actions. In order to study the influence of inequality on the intention of cooperation, the global wealth of the population is fixed to $\epsilon=y\,\epsilon_{{A}}+(1-y)\epsilon_{{B}}$.
Two scenarios will be analyzed:
a completely tolerant resident population (TT) and a population where intolerance has taken over subpopulation A (IT), \ie the scenarios before and after a potential success of an I\textbar TT  invasion.

Figure~\ref{fig:coop} shows the intention of cooperation of individuals from each subpopulation as a function of the Gini coefficient for different compositions of the population.
Interactions are grouped in in-group and out-group behaviors, \ie individuals interacting with members of their own and of the other subpopulation, respectively. The total intention of cooperation for each subpopulation and for the population as a whole is also presented. Results for other parameters values and strategies are displayed in Supplementary Figs.~S1-S6.

From the very definition of the model, it was already known that there exists a general symmetry in the TT scenario, such that results should no be altered by exchanging $y$ for $1-y$ and A for B, as it is confirmed in Fig.~\ref{fig:coop}.
As expected, total cooperation levels of intolerant individuals are lower than those of the tolerant ones, mainly because they never cooperate with members of the other subpopulation. By extension, the aggregate cooperation level in TT scenario is always higher than in the IT scenario.

Results begin to be more interesting when observing the specific behavior between certain subpopulations. In general, intolerant mutant individuals who initially appear in tolerant populations (A$_{M} \rightarrow$A in scenario TT) display significant lower cooperation levels than their tolerant counterparts even with their own kind (A$\rightarrow$A in scenario TT) since they punish them for helping individuals from the other group. However, when these intolerant individuals settle (A$\rightarrow$A in scenario IT), they
become more cooperative with their kind than the tolerant individuals they replace (A$\rightarrow$A in scenario TT).
On the other hand, individuals that remain tolerant reduce their intention of cooperation toward those that become intolerant (B$\rightarrow$A in scenarios TT and IT). This change is amplified when tolerants constitute a disfavored majority ($y<0.5$ and $G>0$) and reduced when they belong to a disfavored minority ($y>0.5$ and $G>0$).

The main differences between strategies \textbf{I} and \textbf{II} are related to the in-group behavior of tolerant minority subpopulations ---A when $y<0.5$ in TT and B when $y>0.5$ in both TT and IT scenarios. The more disfavored these subpopulations are ($G<0$ for A, and $G>0$ for B), the less they intend to cooperate with their kind when following strategy \textbf{II} (linear relationship between intention of cooperation and $G$).
This attitude makes
them more cooperative
with wealthier individuals, even if the latter are intolerant against them (B$\rightarrow$A in scenario IT), than with their own group (B$\rightarrow$B in scenario IT) under disfavored economic conditions (high value of $G$).
On the other hand, individuals following strategy \textbf{I} display a threshold of $G$ from which they begin to increase their intention of cooperation level.
In other words, strategy \textbf{I} tolerant individuals exhibit the highest intention of cooperation toward their own subgroup when their economic conditions are extreme.

For instance, when $y=0.7$ and $G=0.3$, B individuals following strategy \textbf{II} cooperate about three times more with A intolerant individuals than with their own, whereas if they follow strategy \textbf{I} they maintain the maximum level of cooperation regardless of the recipient.
This difference leads to another one: The total intention of cooperation of individuals that play strategy \textbf{II} in tolerant populations is not affected by inequality and is the same no matter if they are in the minority or in the majority. Contrarily, strategy \textbf{I} individuals display higher total intention of cooperation when they belong to the disfavored minority. 
According to how strategies of groups \textbf{I} and \textbf{II} are defined, one can see that the only difference between them is how individuals that recognise themselves as bad should act when encountering other individuals also with bad reputation ($a_{BB}$). Strategists playing \textbf{I} intend to cooperate whereas strategists playing \textbf{II} defect. Moral assessments $m_{BB}(C)$ and $m_{BB}(D)$, respectively, are coherent with these actions assigning a good reputation to the donor after these behaviors. 
Therefore strategy \textbf{I} adopts a relax demeanour when both actors have bad reputation, increasing cooperation when bad economic conditions do not easily allow it, whereas strategy \textbf{II} is more strict punishing defections.

Results for strategy \textbf{III} are shown in Supplementary Figs.~S5 and S6 for completeness. Strategies \textbf{III} display behaviors similar to those from group \textbf{II}. The main quantitative difference is that the former display lower cooperation levels, which is especially significant for tolerant B subpopulation, which more often refuses to help intolerant individuals (IT scenario), except if a high inequality disfavored them. On the other hand, the behavior of this B tolerant subpopulation with its kind is not affected by A becoming intolerant. These attitudes make strategies \textbf{III} more resistant to the invasion of intolerance.

Increasing the global economic stress $\epsilon$ (Supplementary Figs.~S1-S6) does not show a substantial effect on individual behavior beyond reducing cooperation levels ---defection even unintentional usually brings bad reputation. The only exception is mutant intolerant individuals: under some conditions they increase its in-group cooperation level because they see with better eyes their kind not cooperating with the other group, which happens more frequently when global economic stress is higher.

\subsection{Emergence of intolerance}

The next natural step is to analyze under which conditions a small fraction of intolerant
(tolerant) mutants can invade a tolerant (intolerant) resident subpopulation.
Figures~\ref{fig:I} and \ref{fig:II} represent the
parameter regions where each one of the different invasions succeeds in a
$\epsilon_B$--$\epsilon_A$ plane for a given composition of the population $y$. For the sake of clarity, 
the analysis starts
by considering the
I\textbar TT scenario, i.e., intolerant mutants try to invade subpopulation A
of a fully tolerant population. Next it is studied whether tolerance can be restored
provided the first invasion had spread all over subpopulation A (T\textbar IT
scenario), as well as if intolerant mutants can also invade subpopulation B
(IT\textbar I scenario).
The invasion of tolerant mutants of a
fully intolerant population (T\textbar I\,I or I\,I\textbar T) does not succeed for any combination of parameters and then it is not represented.
All the displayed results have been obtained for a benefit-to-cost ratio $b/c=2$; it was already proved \cite{martinez:2014} that reducing this ratio has the effect of favoring the general spread of intolerance and it does not offer any significant new finding for the present work.

Consistent with previous findings \cite{martinez:2014}, one observes that
minorities are susceptible to intolerant outbreaks when they are under high
economic stress $\epsilon_A$ (blue region in Figs.~\ref{fig:I} and
\ref{fig:II} for low values of $y$). But these figures also illustrate the
effect of economic inequality. For a given composition of the population $y$, moving along the lines
with constant $\epsilon = y\,\epsilon_{{A}}+(1-y)\epsilon_{{B}}$ in the
$\epsilon_B$--$\epsilon_A$ plane (grey lines in Figures~\ref{fig:I} and \ref{fig:II}) keeps
constant the total population wealth while at the same time changes the Gini
coefficient $G$ ---which will then be proportional to the difference
$\epsilon_{{B}}-\epsilon_{{A}}$ according to Eq.~\ref{eq:Gini}. Above the diagonal $\epsilon_B=\epsilon_A$, $G<0$ (A is disfavored), whereas below the diagonal $G>0$ (A is favored).

The general trend
is that egalitarian ($G=0$) fully tolerant populations are able to repel intolerance become sensitive to its invasion  by increasing inequality.
On the other hand, a
population for which intolerance can spread when $G=0$ (blue region in Figs.~\ref{fig:I} and \ref{fig:II}) can be made
intolerance-proof by introducing some economic inequality ($G\ne 0$) in favor of
the group targeted by intolerance.
This holds up to a certain
level of global economic stress in the case of strategy \textbf{II} minorities; beyond that, intolerance invades regardless
of the value of $G$.

Likewise, full tolerance can be restored in a population IT  (red region of Figs.~\ref{fig:I}
and \ref{fig:II}) by modifying
$G$ in favor of the intolerant minority 
that is susceptible to be invaded by tolerant individuals.
In general, the lower
the global economic stress the less inequality is needed
for tolerance invasion to succeed.
However this trend reverses if the
intolerants constitute the majority; in this scenario the lower the global economic stress the
more inequality is needed to restore tolerance (and below a threshold tolerance
can no longer be restored).
Purple regions (overlaps of blue and red regions) are combinations of
parameters for which I\textbar TT and T\textbar IT invasions can both occur.
This means that neither TT or IT populations are stable, and the equilibrium
will most likely consist of a mixture of tolerant and intolerant individuals in
the same subpopulation. This effect can only be observed if the invaded
population is a minority and if inequality favors them. In general, though,
intolerance outbreaks will invade irreversibly.

Finally, Figs.~\ref{fig:I} and \ref{fig:II} also show that,
once intolerance has settled in one group,
it is almost unavoidable that the whole population becomes completely intolerant regardless of the global economic stress and the Gini coefficient, 
especially if the intolerant group constitutes the majority. Thus, 
populations are at high risk of
becoming fully intolerant in
regions where  I\textbar TT and  IT\textbar I overlaps
If this happens there is no way back:
A fully intolerant population can never be reverted to tolerance by changing the
economic parameters.

\subsection{Redistribution of wealth}

\begin{figure*}
	\includegraphics[width=6in]{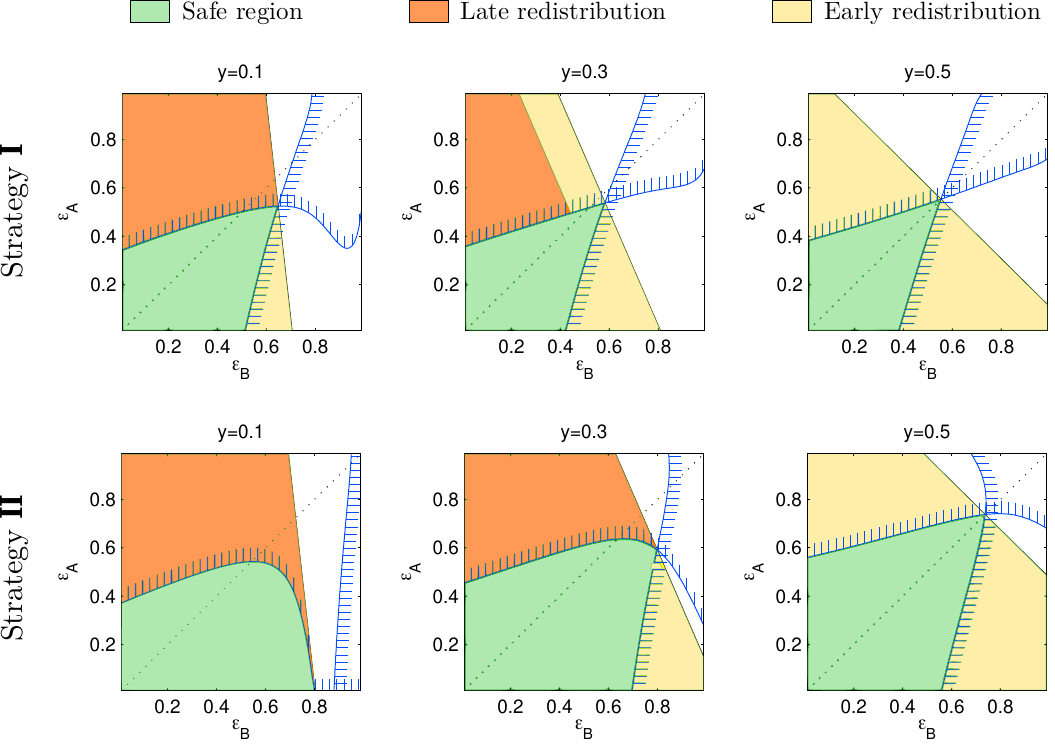}
	\caption{{Redistribution of wealth to prevent intolerance}. Areas in the $\epsilon_A$--$\epsilon_B$ plane, for different compositions of the population $y$ and strategies \textbf{I} and \textbf{II}, where (\textit{i}) intolerance cannot invade any of the
		subpopulations (\textit{safe region}), (\textit{ii}) a redistribution of the wealth can
		restore tolerance in the smallest subpopulation once intolerance has already invaded it (\textit{late redistribution}),
		(\textit{iii}) a redistribution of wealth before an outbreak of intolerance can take the whole population
		to the safe zone (\textit{early} and \textit{late redistribution}). Blue lines mark the boundaries for the success of the invasions
		I\textbar TT and T\textbar IT.  Dotted lines correspond to $\epsilon_A=\epsilon_B$. Scenarios with $y> 0.5$ are equivalent to those shown with $y< 0.5$ by exchanging $\epsilon_A$ for $\epsilon_B$, and $y$ for $1-y$. Benefit-to-cost ratio $b/c=2$.}
	\label{fig:bi}
\end{figure*}

The picture that has been presented shows that economic inequality can be an enhancer
of intolerance, even more so than the global economic stress. Populations that
would resist intolerance outbreaks can easily fall into intolerant attitudes if
wealth is unequally distributed. It has already been shown that modifying inequality affects the ability of tolerance and intolerance to emerge, so then it makes sense to ask whether a redistribution of wealth can prevent intolerance from succeeding.

Assuming a fully tolerant population, we can gather together the areas where I\textbar TT and TT\textbar I invasions succeed, \ie blue regions of Figs.~\ref{fig:I} and \ref{fig:II} corresponding to $y$ (subpopulation A)
and $1-y$ (subpopulation B), and delimit a region of the plane $\epsilon_A$--$\epsilon_B$
where the population is intolerance-proof (\textit{safe region}).
This region is displayed in green in Figure~\ref{fig:bi}.
If the distribution of wealth in a
population has a distribution within the safe region, tolerance is stable.

Redistributing wealth amounts to changing $\epsilon_A$ and $\epsilon_B$ along the line
$y\epsilon_A+(1-y)\epsilon_B$ maintaining $\epsilon$ constant. A wealth redistribution policy can only be successful if this
line crosses the safe region. Orange and yellow areas in Fig.~\ref{fig:bi}
satisfy this condition. Outside those areas, redistributing wealth is an useless
policy against outbreaks of intolerance. Nevertheless, the effect in
these two areas is different.

Within the \textit{late redistribution} region (orange in Fig.~\ref{fig:bi}), an invasion I\textbar TT can occur and transform the
population into IT. However, full tolerance can be restored in one or
two steps. First, reversing wealth inequality so as to bring the population
into one of the zones where
T\textbar IT invasion is possible (red zones in Figs.~\ref{fig:I} and \ref{fig:II}). Second, once full tolerance is restored,
and if the population is not already in the safe region,
wealth must be redistributed again to bring the population to the safe
region.

On the other hand, within  the \textit{early redistribution} region (yellow in Fig.~\ref{fig:bi})
wealth redistribution is only useful as a prevention
policy: it has to be implemented \emph{before} an
intolerance outbreak takes place. Otherwise, since the population cannot be
moved into one of the red zones of Figs.~\ref{fig:I} and \ref{fig:II}, once
one of the groups becomes intolerant, chances are that the tolerant
group eventually becomes intolerant as well (population falls in
regions where IT\textbar I or I\textbar TI invasions can prosper). The only
possibility to revert the population to TT is to move it into a region where
T\textbar IT (TI\textbar T) and  IT\textbar I (I\textbar TI) areas overlap. However, since both invasions can occur, there is no guarantee that the policy succeeds.
If the population becomes completely intolerant, tolerance will never be restored even if the
economic situation improves.

\section{Discussion}

The connection between global economic stress and outbursts of intolerance is well
documented \cite{dollard:1939, hovland:1940, lipset1959democracy, levin:1982, persell2001civil,  gibson:2002, svallfors2006moral}, and the
present model of indirect reciprocity yields results in
agreement with these observations.
However, the aim of this work is to study the effect of economic inequality in the evolutionary emergence of intolerance.
By letting subpopulations to undergo different levels of economic stress, inequality has been proved as a powerful enhancer of intolerance, making it easier for intolerant individuals to prosper even when global economic conditions are not too harsh.
Additionally, discriminatory behaviors can easily escalate without the presence of new intolerant mutants. For instance, if the disfavored tolerant individuals constitute a majority, they withdraw their support from the wealthy intolerant minority simply by following their intrinsic strategy. 

Results show how inequality and intolerance, despite having been defined as a type of out-group discrimination, have an important impact not only on attitudes against other group members but also toward their own kind. These behaviors are different for different evolutionarily stable strategies, revealing a diverse spectrum of possibilities to tackle intolerance under unequal populations.  
Some strategists intend to behave as cooperatively as possible with other tolerant individuals of their kind when they belong to the disfavored minority. This attitude can be interpreted as a sympathetic behavior that allows them to avoid the spread of bad reputations originated by their inability to cooperate  when resources are scarce for them. In return, they pay the price of reducing their punishment capacity against individuals that have fairly acquired bad reputation.
Other strategies, on the other hand, promote a more rigid behavior by never cooperating with individuals who have earned a bad reputation regardless of their own reputation and the disadvantaged conditions in which they may find themselves. However, it has the counterintuitive consequence that disfavored individuals help more frequently to wealthier individuals, even if they display intolerant attitudes against them, than to those from their own kind when they are in minority. Some experiments have shown this kind of attitudes\cite{hackel2018propagation}, at least when social identity is not shared \cite{gangadharan2019impact}, which can propagate economic inequality.
These differences among strategies bear some parallelism with the aspiration level of  win-stay, lose-shift strategies in direct reciprocity \cite{nowak:1995, martinez:2012}: some donors display a lower aspiration level, cooperating not only when the recipient enjoys a good reputation, but also when both the recipient and herself own bad reputations, obtaining a good reputation in return; other donors hold higher aspiration levels and cooperate only if the recipient displays a good reputation.
On the other hand, the capacity of individuals to punish bad behaviors is generally limited by their economic power. Under limited resources they are not able to efficiently implement, even when they intend to, different attitudes toward what they consider good and bad behaviors.  It has been previously shown that the lack of resources, such as money or time, generally leads to poorer decisions due to the cognitive load imposed by poverty \cite{mani2013poverty}.
Despite the fact that
intolerance outbreaks more easily  in disfavored minorities,
results have also shown that exceptionally a wealthy small minority within an unequal population can oscillate between tolerant and intolerant behaviors (or a mixed equilibrium of them). 

The present study was able to delimit regions of inequality where intolerant
behaviors cannot spread. More importantly, results have
also shown regions where inequality is so high that it favors the
appearance of intolerance, but where a proper redistribution of wealth may
prevent that from happening. In order to become a viable solution, this redistribution has to fulfil two key requirements.
First, it needs to favor equality: the emergence of intolerance can be avoided favoring the group that is susceptible of this invasion but if it is done at the expenses of increasing inequality, intolerance eventually may invade the disfavored group. Therefore, increasing inequality as a mean to avoid intolerance only has sense if the latter group owns specific features (external to the model) that make it resilience to intolerance. Secondly, the redistribution of wealth has to be applied
\emph{before} intolerance emerges. Otherwise it may not be as effective, or
even be plainly useless.

The model used in this work is affected by some simplifications to make the fundamental analysis feasible. For instance, the redistribution of wealth was implemented at no cost, and therefore, the results of the model are to be taken as the best-case scenario of what such a policy is able to achieve.
There exist a number of other potential aspects beyond the scope of this work that could be studied increasing the level of complexity of the model \cite{santos2021complexity}: the impact of individuals' mobility across groups \cite{fu:2012, whitaker2018indirect, zhang2021effects, oliveira2022group}, the effect of a higher order reputation system \cite{murase2022social}, the importance of the quality of the information perceived in the private assessments \cite{hilbe2018indirect}, the possibility of signalling or communication \cite{martinez2020signalling}, the introduction of apology mechanisms \cite{martinez2015apology}, the influence of different scales in finite size populations and network topologies \cite{karimi2018homophily, hu2021adaptive, dimaggio2012network}.
Handling these elements, together with other externalities such as
the global economic power or the benefit-to-cost ratio, may help to limit the appearance of intolerance in situations where the redistribution of wealth alone is not enough.

The findings of this research  are especially relevant given the rise of inequality in modern societies \cite{firebaugh2000trend, goesling2001changing,  fischer2006century, piketty2018distributional, Alvaredo+2018, blanchet2019unequal} increasingly reliant on public reputation systems, which are not able to reduce initial inequalities \cite{kas2022role}.
Moreover, the recent COVID-19 global pandemic has boosted the situation, leaving the most vulnerable groups in an even more precarious situation \cite{chancel2021world}.
Inequality, without appropriate policies to curb it, escalates \cite{browman2019economic}, and brings significant dangers \cite{jetten2017social, jetten2019social}: It is detrimental to the economy \cite{bowles2012new} and a source of socio-political instability \cite{russett1964inequality, huntington1968political, alesina1996income, perotti1996growth}, being a threat to democracy \cite{andersen2012support} and even the seed of civil wars \cite{cederman2013inequality}.
There exist clear indications that economic inequality has a direct connection with the emergence of intolerance in many current phenomena, such as Brexit or the rise of far-right populism \cite{inglehart2016trump, meleady2017examining}, which strategically use problems derived from inequality to limit the redistribution of wealth and maintain a spiral of inequality \cite{jay2019economic}.
Other policies designed to directly protect the rights of minority groups (\eg their distinctness) could yield the opposite effect they intend to, at least if economic inequality is preserved \cite{sniderman2009ways}.
On the other hand,
experiments show that redistribution of wealth policies would be accepted by a significant part of society, especially if they are accompanied by a more clear perception of the existence of economic inequality \cite{garcia2020perceiving, chisadza2021group,kirkland2021economic} and are fairly and efficiently implemented \cite{koster2022human}.
All these evidences
are consistent with
the results of this work: the importance of implementing wealth redistribution policies before intolerance permeates part of the population.

\section*{Supplementary Material}
 Supplementary Figures represent the intention of cooperation as a function of the Gini coefficient for additional model parameter values.

\section*{Author Declarations}
The author has no conflicts to disclose.

\begin{acknowledgements}
	The author thanks José A. Cuesta and The Anh Han for helpful comments and suggestions.
	This work was 
	supported by the Ministry of Science and Innovation (Spain) under the project PID2021-122711NB-C21.
\end{acknowledgements}

\appendix*

\section{Reputation equilibria under different scenarios }

As it was stated in the main text, the interactions with recipients of the same and of the opposite group, introduced in Eqs.~\ref{eq:evol_xA_gen}--\ref{eq:evol_xMB_gen}, were split into $T_{i,j}^{\Lambda_l\Lambda_m}$ and $F_{i,j}^{\Lambda_l\Lambda_m}$,
respectively. Fractions $T_{i,j}^{\Lambda_l\Lambda_m}$ are common for all invasion scenarios and can
be obtained as
\begin{widetext}
\begin{eqnarray}
	T_{i,A}^{\Lambda_A\Lambda_M} =&& \sum_{\alpha_A\alpha_M\beta_A\beta_M} x_i^{\alpha_A*\alpha_M} x_A^{\beta_A*\beta_M}  Q^{\Lambda_A\Lambda_M}(\alpha_A\beta_A,\alpha_M\beta_M|\alpha_i\beta_i) \\
	T_{i,B}^{\Lambda_B\Lambda_M} =&& \sum_{\alpha_i\alpha_M\beta_B\beta_M} x_i^{*\alpha_B\alpha_B} x_B^{*\beta_B\beta_M}  Q^{\Lambda_B\Lambda_M}(\alpha_B\beta_B,\alpha_M\beta_M|\alpha_i\beta_i) \\
	T_{i,i}^{\Lambda_A\Lambda_B} =&& \sum_{\alpha_A\alpha_B\beta_A\beta_B} x_i^{\alpha_A\alpha_B*} x_i^{\beta_A\beta_B*}  Q^{\Lambda_A\Lambda_B}(\alpha_A\beta_A,\alpha_B\beta_B|\alpha_B\beta_B) \\
			Q  ^{\Lambda_l\Lambda_m} (\alpha_l\beta_l,\alpha_m\beta_m|\alpha_i\beta_i) &&=  (1-\epsilon_i)\,  \delta \big( \Lambda_l, m_{\alpha_l\beta_l}(a_{\alpha_i\beta_i}) \big) 
	\delta \big( \Lambda_m, m_{\alpha_m\beta_m}(a_{\alpha_i\beta_i})\big) 
	+\, \epsilon_i\, \delta \big( \Lambda_l, m_{\alpha_l\beta_l}(D) \big) 
\end{eqnarray}
\end{widetext}
where $Q^{\Lambda_l\Lambda_m}(\alpha_l\beta_l,\alpha_m\beta_m|\alpha_i\beta_i)$ is the probability
that an $i$-type donor with reputation $(\alpha_l,\alpha_m)$ acting on a recipient
with reputation $(\beta_l,\beta_m)$, is assigned a reputation $(\Lambda_l,\Lambda_m)$ by $l$-individuals and $m$-individuals.
On the other hand, fractions $F_{i,j}^{\Lambda_l\Lambda_m}$ depend on the specific scenario and are described in the following sections.

\subsection{Auxiliary functions}
\label{sec:apen_aux}

Firstly, some auxiliary functions and the process to calculate reputations in homogeneous populations is presented. The main auxiliary function that will be used in the later mathematical process is
\begin{equation}
	U(x_i,x_j) =\sum_{\alpha \beta} \chi_\alpha(x_i) \chi_\beta(x_j) R_i^{\alpha \beta},
	\label{eq:fU}
\end{equation}
where
\begin{equation}
	R_i^{\alpha \beta} = (1-\epsilon_i) m_{\alpha \beta}(a_{\alpha \beta}) + \epsilon_i m_{\alpha \beta}(D)
\end{equation}
represents the probability that a $i$-type donor  with reputation $\alpha$ performing the action
$a_{\alpha \beta}$ on a recipient with reputation $\beta$ is considered good according
to the moral assessment $ m_{\alpha \beta}(a)$.

In a homogeneous population---where all its members belong to the same $i$-type---the
dynamics of the fraction of good individuals $g_i$ follows the differential
equation
\begin{equation}
	\frac{dg_i}{dt}=U(g_i,g_i)-g_i.
\end{equation}
In equilibrium $g_i=U(g_i,g_i)$, which is a quadratic equation with a unique stable
solution $0\le g_i\le 1$.
A continuous perturbation of the
above equation will be also consider:
\begin{equation}
	x_i={r}\,U(x_i,x_i)+1-{r},
\end{equation}
and denote $x_i=p_i({r})$ its stable solution in the interval $[0,1]$.  The solution of the unperturbed equation is recovered with
$p_i(1)=g_i$.

Another function that will be used is ${q_i}({r},{s})$, which correspond to the solution of the linear equation
$x_i={r}\,U(x_i,s)+1-{r}$, $0\le {r},{s}\le 1$:
\begin{equation}
		{q_i}({r},{s}) = \frac{1+{r}\lc {s}R_i^{BG} + (1-{s})R_i^{BB}
			-1\rc }{1-{r}\lc {s}R_i^{GG}-(1-{s})R_i^{GB}+kP_{BG} + (1-{s})R_i^{BB}\rc} .
\end{equation}

\subsection{$I|TT$ scenario}

Since residents
do not distinguish classes, the resident population acts as a homogeneous
population. Therefore $x_i^{*\Lambda_B\Lambda_M}=x_i^{\Lambda_A*\Lambda_M}$,
$\eta_{B,A}=\eta_{A,B}=\eta_{A,A}$, $\eta_{B,M}=\eta_{A,M}$, and
$\eta_{M,B}=0$. The fraction of good residents is $g_A^{A}=p_A(1)$ and $g_B^{A}=p_B(1)$, and accordingly,
\begin{eqnarray}
		x_A^{G*B}=g_A^{A}-x_A^{G*G},
		\label{eq:x1GB} \\
		x_A^{B*G}=1-g_A^{A}-x_A^{B*B}.
\end{eqnarray}
Therefore the fraction $F^{\Lambda_A\Lambda_M}_{A,B}$ introduced in Eq.~\ref{eq:evol_xA_gen} can be expressed in
this scenario as
\begin{eqnarray}
		F^{\Lambda_A\Lambda_M}_{A,B}&&=\sum_{\alpha \beta} \chi_{\alpha}(g^A_A) \chi_{\beta}(g^A_B)	P_A^{\Lambda_A\Lambda_M}(\alpha \beta), \\
		P_A^{\Lambda_A\Lambda_M}(\alpha \beta)\ &&= \ (1-\epsilon_A)\,  \delta  \lp
		\Lambda_A, m_{\alpha \beta}(a_{\alpha \beta}) \rp  \delta \lp \Lambda_M,  \tilde{a}_{\alpha \beta}\rp \ \nonumber\\ 
		&& +\ \epsilon_A\, \delta \lp \Lambda_A, m_{\alpha \beta}(D)\rp \delta \lp \Lambda_M, G \rp,
\end{eqnarray}
where  $\tilde{a}_{\alpha \beta}$ stands for the opposite action to $a_{\alpha \beta}$. Once we numerically solve the two Eqs.~\ref{eq:evol_xA_gen},
Eq.~\ref{eq:evol_xM_gen}, which corresponds to the intolerant mutants, can be
solved analytically by taking into account that
\begin{eqnarray}
		F^{\Lambda_A\Lambda_M}_{M,B}=\sum_{\alpha \beta} \chi_{\alpha}(g^A_M) \chi_{\beta}(g^A_B)
		P_M^{\Lambda_A\Lambda_M}(\alpha \beta), \\ P_M^{\Lambda_A\Lambda_M}(\alpha \beta)\ =\  \delta \lp \Lambda_A,
		m_{\alpha \beta}(D) \rp \delta \lp \Lambda_M, G \rp.
\end{eqnarray}
If at equilibrium, Eq.~\ref{eq:evol_xM_gen} is summed over $\Lambda_A$ and $g_M^{M}$ is given by ${q_A}(g_A^{M},y)$. Then the set of
Eqs.~\ref{eq:evol_xM_gen} reduces to just two equations.

\subsection{$T|IT$ scenario}

Since mutants and B-residents are both tolerant, they will judge every player equally
and then $x_i^{\Lambda_A\Lambda_B*}=x_i^{\Lambda_A*\Lambda_M}$. On the other hand, A-residents are intolerant
($\eta_{A,B}=0$) and their dynamics is described by
Eq.~\ref{eq:evol_xA_gen}, where
\begin{eqnarray}
		F^{\Lambda_A\Lambda_M}_{A,B}=\sum_{\alpha \beta} \chi_{\alpha}(g^M_A) \chi_{\beta}(g^M_B)
		P_A^{\Lambda_A\Lambda_M}(\alpha \beta), \\ P_A^{\Lambda_A\Lambda_M}(\alpha \beta)\ =\  \delta \lp \Lambda_M,
		m_{\alpha \beta}(D) \rp \delta \lp \Lambda_A, G \rp.
\end{eqnarray}
Summing Eq.~\ref{eq:evol_xA_gen} over $\Lambda_M$ in equilibrium one obtains that
$g_A^{A}=p_A(y)$. In this scenario it is also necessary to calculate $g_B^{M}$.
Since $B$-mutant's judgements are the same as those of B-residents,
$g_B^{M}$ is given at equilibrium by
\begin{equation}
	g_B^{M}\ =\ (1-y) U(g_B^{M},g_B^{M}) + y U(g_B^{M},g_A^{M}).
	\label{eq:equil_xB_TIT}
\end{equation}
Therefore,  a system of three equations ---two from
Eqs.~\ref{eq:evol_xA_gen} plus Eq.~\ref{eq:equil_xB_TIT}--- needs to be solved numerically.
After that, the set of Eqs.~\ref{eq:evol_xM_gen} with
\begin{eqnarray}
		F^{\Lambda_A\Lambda_M}_{M,B}&&=\sum_{\alpha \beta} \chi_{\alpha}(g^M_M)
		\chi_{\beta}(g^B_M) P_M^{\Lambda_A\Lambda_M}(\alpha \beta), \\
		P_M^{\Lambda_A\Lambda_M}(\alpha \beta)\ &&=\ (1-\epsilon_A)\, \delta \lp
		\Lambda_M, m_{\alpha \beta}(a_{\alpha \beta}) \rp  \delta \lp \Lambda_A,  \tilde{a}_{\alpha \beta}\rp \ \nonumber\\
		&&+\ \epsilon_A\, \delta \lp \Lambda_M, m_{\alpha \beta}(D)\rp  \delta \lp \Lambda_A, G \rp,
\end{eqnarray}
can be solved analytically, taking into account that summing
over $\Lambda_A$ at equilibrium yields $g_M^{M}={q_A}\lp y g_A^{M}+(1-y)
g_B^{M} , 1 \rp$.

\subsection{$IT|I$ scenario}

In this scenario, mutants belongs to subpopulation B and intolerant toward A-residents, who are intolerant in turn toward B-type individuals. Hence
$\eta_{A,B}=\eta_{A,M}=\eta_{M,A}=0$.

In order to calculate $g_A^B$ we first need to compute $x_A^{\Lambda_A\Lambda_B*}$
through Eq.~\ref{eq:evol_xA1_gen}, where
\begin{eqnarray}
		F^{\Lambda_A\Lambda_B}_{A,B}=\sum_{\alpha \beta} \chi_{\alpha}(g^B_A) \chi_{\beta}(g^B_B)
		P_A^{\Lambda_A\Lambda_B}(\alpha \beta), \\ P_A^{\Lambda_A\Lambda_B}(\alpha \beta)\ =\  \delta \lp \Lambda_B,
		m_{\alpha \beta}(D) \rp \delta \lp \Lambda_A, G \rp,
\end{eqnarray}
and summing over $\Lambda_B$ at equilibrium one obtains that $g_A^{A}= p_A(y)$. In
this way, we reduced the set of Eq.~\ref{eq:evol_xA1_gen} to just two
equations.

On the other hand, the dynamics of the tolerant individuals is given by
Eq.~\ref{eq:evol_xB_gen}, with
\begin{eqnarray}
		F^{\Lambda_B\Lambda_M}_{B,A}&&=\sum_{\alpha \beta} \chi_{\alpha}(g^B_B) \chi_{\beta}(g^B_A)
		P_B^{\Lambda_B\Lambda_M}(\alpha \beta), \\
		P_B^{\Lambda_B\Lambda_M}(\alpha \beta)\ &&= \ (1-\epsilon_B)\, \delta  \lp
		\Lambda_B, m_{\alpha \beta}(a_{\alpha \beta}) \rp \delta \lp \Lambda_M,  \tilde{a}_{\alpha \beta}\rp \ \nonumber\\
		&&+\ \epsilon_B\, \delta \lp \Lambda_B, m_{\alpha \beta}(D)\rp  \delta \lp \Lambda_M, G \rp.
\end{eqnarray}
Taking into account that $x_B^{*AG}=1-x_B^{*GG}-x_B^{*GB}-x_B^{*BB}$, five coupled equations needs to be solved, namely, three from Eqs.~\ref{eq:evol_xA_gen} and
two from Eqs.~\ref{eq:evol_xB1_gen}.

The dynamics of the mutants is solved from Eq.~\ref{eq:evol_xMB_gen}, where
\begin{eqnarray}
		F^{\Lambda_B\Lambda_M}_{M,A}=\sum_{\alpha \beta} \chi_{\alpha}(g^B_M) \chi_{\beta}(g^B_A)
		P_M^{\Lambda_B\Lambda_M}(\alpha \beta), \\ P_M^{\Lambda_B\Lambda_M}(\alpha \beta)\ =\  \delta \lp \Lambda_B,
		m_{\alpha \beta}(D) \rp \ \delta \lp \Lambda_M, G \rp.
\end{eqnarray}
Summing over $\Lambda_B$ at equilibrium yields $g_M^{M}={q_B}(g_B^{M},1-y)$, and
reduced the last set of equations to only two linear.

\subsection{$T|II$ scenario}

In this last scenario a small fraction of tolerant mutants, who belong to subpopulation A, tries to invade a
completely intolerant population. Therefore $\eta_{A,B}=\eta_{B,A}=\eta_{B,M}=0$.
Similar analysis can be done for B-type mutants and $I|TI$ invasion exchanging A for B and $y$ for $1-y$.

The dynamics for the intolerant A-residents is described by
Eq.~\ref{eq:evol_xA_gen} with
\begin{eqnarray}
		F^{\Lambda_A\Lambda_M}_{A,B}=\sum_{\alpha \beta} \chi_{\alpha}(g^M_A) \chi_{\beta}(g^M_B)
		P_A^{\Lambda_A\Lambda_M}(\alpha \beta), \\ P_A^{\Lambda_A\Lambda_M}(\alpha \beta)\ =\  \delta \lp \Lambda_M,
		m_{\alpha \beta}(D) \rp \delta \lp \Lambda_A, G \rp.
\end{eqnarray}
Summing over $\Lambda_M$ at equilibrium yields $g_A^{A} = p(y)$.

Computing first $x_B^{*\Lambda_B\Lambda_M}$ is necessary in order to calculate $g_B^{M}$. Equation~\ref{eq:evol_xB_gen} describes the dynamics for these fractions, where
\begin{eqnarray}
		F^{\Lambda_B\Lambda_M}_{B,A}=\sum_{\alpha \beta} \chi_{\alpha}(g^M_B) \chi_{\beta}(g^M_A)
		P_B^{\Lambda_B\Lambda_M}(\alpha \beta), \\ P_B^{\Lambda_B\Lambda_M}(\alpha \beta)\ =\  \delta \lp \Lambda_M,
		m_{\alpha \beta}(D) \rp \delta \lp \Lambda_B, G \rp,
\end{eqnarray}
and summing again over $\Lambda_M$ one obtains that $g_B^{B} = p(1-y)$. The four resulting coupled equations ---two from Eqs.~\ref{eq:evol_xB1_gen} and two from Eq.~\ref{eq:evol_xB_gen}--- need to be solved numerically.

On the other hand, the dynamics of the tolerant mutants is given by
Eq.~\ref{eq:evol_xB_gen}, where
\begin{eqnarray}
		F^{\Lambda_A\Lambda_M}_{M,B}&&=\sum_{\alpha \beta} \chi_{\alpha}(g^M_M) \chi_{\beta}(g^M_B)
		P_M^{\Lambda_A\Lambda_M}(\alpha \beta), \\
		P_M^{\Lambda_A\Lambda_M}(\alpha \beta) && =\ (1-\epsilon_A)\, \delta  \lp
		\Lambda_M, m_{\alpha \beta}(a_{\alpha \beta}) \rp  \delta \lp \Lambda_A, \tilde{a}_{\alpha \beta}\rp \nonumber\\
		&&+ \epsilon_A\, \delta \lp \Lambda_M, m_{\alpha \beta}(D)\rp \delta \lp \Lambda_A, G \rp.
\end{eqnarray}
This set of equations can be reduced to just two because summing over $\Lambda_A$ at
equilibrium yields $g_M^{M}={q_i}( yg_A^{M}+(1-y)g_B^{M},1)$.


\begin{thebibliography}{104}%
	\makeatletter
	\providecommand \@ifxundefined [1]{%
		\@ifx{#1\undefined}
	}%
	\providecommand \@ifnum [1]{%
		\ifnum #1\expandafter \@firstoftwo
		\else \expandafter \@secondoftwo
		\fi
	}%
	\providecommand \@ifx [1]{%
		\ifx #1\expandafter \@firstoftwo
		\else \expandafter \@secondoftwo
		\fi
	}%
	\providecommand \natexlab [1]{#1}%
	\providecommand \enquote  [1]{``#1''}%
	\providecommand \bibnamefont  [1]{#1}%
	\providecommand \bibfnamefont [1]{#1}%
	\providecommand \citenamefont [1]{#1}%
	\providecommand \href@noop [0]{\@secondoftwo}%
	\providecommand \href [0]{\begingroup \@sanitize@url \@href}%
	\providecommand \@href[1]{\@@startlink{#1}\@@href}%
	\providecommand \@@href[1]{\endgroup#1\@@endlink}%
	\providecommand \@sanitize@url [0]{\catcode `\\12\catcode `\$12\catcode
		`\&12\catcode `\#12\catcode `\^12\catcode `\_12\catcode `\%12\relax}%
	\providecommand \@@startlink[1]{}%
	\providecommand \@@endlink[0]{}%
	\providecommand \url  [0]{\begingroup\@sanitize@url \@url }%
	\providecommand \@url [1]{\endgroup\@href {#1}{\urlprefix }}%
	\providecommand \urlprefix  [0]{URL }%
	\providecommand \Eprint [0]{\href }%
	\providecommand \doibase [0]{http://dx.doi.org/}%
	\providecommand \selectlanguage [0]{\@gobble}%
	\providecommand \bibinfo  [0]{\@secondoftwo}%
	\providecommand \bibfield  [0]{\@secondoftwo}%
	\providecommand \translation [1]{[#1]}%
	\providecommand \BibitemOpen [0]{}%
	\providecommand \bibitemStop [0]{}%
	\providecommand \bibitemNoStop [0]{.\EOS\space}%
	\providecommand \EOS [0]{\spacefactor3000\relax}%
	\providecommand \BibitemShut  [1]{\csname bibitem#1\endcsname}%
	\let\auto@bib@innerbib\@empty
	\bibitem [{\citenamefont {Mattison}\ \emph {et~al.}(2016)\citenamefont
		{Mattison}, \citenamefont {Smith}, \citenamefont {Shenk},\ and\ \citenamefont
		{Cochrane}}]{mattison2016evolution}%
	\BibitemOpen
	\bibfield  {author} {\bibinfo {author} {\bibfnamefont {S.~M.}\ \bibnamefont
			{Mattison}}, \bibinfo {author} {\bibfnamefont {E.~A.}\ \bibnamefont {Smith}},
		\bibinfo {author} {\bibfnamefont {M.~K.}\ \bibnamefont {Shenk}}, \ and\
		\bibinfo {author} {\bibfnamefont {E.~E.}\ \bibnamefont {Cochrane}},\
	}\bibfield  {title} {\enquote {\bibinfo {title} {The evolution of
				inequality},}\ }\href@noop {} {\bibfield  {journal} {\bibinfo  {journal}
			{Evolutionary Anthropology: Issues, News, and Reviews}\ }\textbf {\bibinfo
			{volume} {25}},\ \bibinfo {pages} {184--199} (\bibinfo {year}
		{2016})}\BibitemShut {NoStop}%
	\bibitem [{\citenamefont {Alvaredo}(2018)}]{Alvaredo+2018}%
	\BibitemOpen
	\bibfield  {author} {\bibinfo {author} {\bibfnamefont {F.}~\bibnamefont
			{Alvaredo}},\ }\href@noop {} {\emph {\bibinfo {title} {The World Inequality
				Report: 2018}}}\ (\bibinfo  {publisher} {Harvard University Press},\ \bibinfo
	{year} {2018})\BibitemShut {NoStop}%
	\bibitem [{\citenamefont {Mazur}(2000)}]{mazur2000labor}%
	\BibitemOpen
	\bibfield  {author} {\bibinfo {author} {\bibfnamefont {J.}~\bibnamefont
			{Mazur}},\ }\bibfield  {title} {\enquote {\bibinfo {title} {Labor's new
				internationalism},}\ }\href@noop {} {\bibfield  {journal} {\bibinfo
			{journal} {Foreign Aff.}\ }\textbf {\bibinfo {volume} {79}},\ \bibinfo
		{pages} {79} (\bibinfo {year} {2000})}\BibitemShut {NoStop}%
	\bibitem [{\citenamefont {Firebaugh}(2000)}]{firebaugh2000trend}%
	\BibitemOpen
	\bibfield  {author} {\bibinfo {author} {\bibfnamefont {G.}~\bibnamefont
			{Firebaugh}},\ }\bibfield  {title} {\enquote {\bibinfo {title} {The trend in
				between-nation income inequality},}\ }\href@noop {} {\bibfield  {journal}
		{\bibinfo  {journal} {Annual Review of Sociology}\ ,\ \bibinfo {pages}
			{323--339}} (\bibinfo {year} {2000})}\BibitemShut {NoStop}%
	\bibitem [{\citenamefont {Goesling}(2001)}]{goesling2001changing}%
	\BibitemOpen
	\bibfield  {author} {\bibinfo {author} {\bibfnamefont {B.}~\bibnamefont
			{Goesling}},\ }\bibfield  {title} {\enquote {\bibinfo {title} {Changing
				income inequalities within and between nations: New evidence},}\ }\href@noop
	{} {\bibfield  {journal} {\bibinfo  {journal} {American Sociological Review}\
			,\ \bibinfo {pages} {745--761}} (\bibinfo {year} {2001})}\BibitemShut
	{NoStop}%
	\bibitem [{\citenamefont {Fischer}\ and\ \citenamefont
		{Hout}(2006)}]{fischer2006century}%
	\BibitemOpen
	\bibfield  {author} {\bibinfo {author} {\bibfnamefont {C.~S.}\ \bibnamefont
			{Fischer}}\ and\ \bibinfo {author} {\bibfnamefont {M.}~\bibnamefont {Hout}},\
	}\href@noop {} {\emph {\bibinfo {title} {Century of difference: How America
				changed in the last one hundred years}}}\ (\bibinfo  {publisher} {Russell
		Sage Foundation},\ \bibinfo {year} {2006})\BibitemShut {NoStop}%
	\bibitem [{\citenamefont {Piketty}, \citenamefont {Saez},\ and\ \citenamefont
		{Zucman}(2018)}]{piketty2018distributional}%
	\BibitemOpen
	\bibfield  {author} {\bibinfo {author} {\bibfnamefont {T.}~\bibnamefont
			{Piketty}}, \bibinfo {author} {\bibfnamefont {E.}~\bibnamefont {Saez}}, \
		and\ \bibinfo {author} {\bibfnamefont {G.}~\bibnamefont {Zucman}},\
	}\bibfield  {title} {\enquote {\bibinfo {title} {Distributional national
				accounts: methods and estimates for the united states},}\ }\href@noop {}
	{\bibfield  {journal} {\bibinfo  {journal} {The Quarterly Journal of
				Economics}\ }\textbf {\bibinfo {volume} {133}},\ \bibinfo {pages} {553--609}
		(\bibinfo {year} {2018})}\BibitemShut {NoStop}%
	\bibitem [{\citenamefont {Blanchet}\ \emph {et~al.}(2019)\citenamefont
		{Blanchet}, \citenamefont {Chancel}, \citenamefont {Gethin} \emph
		{et~al.}}]{blanchet2019unequal}%
	\BibitemOpen
	\bibfield  {author} {\bibinfo {author} {\bibfnamefont {T.}~\bibnamefont
			{Blanchet}}, \bibinfo {author} {\bibfnamefont {L.}~\bibnamefont {Chancel}},
		\bibinfo {author} {\bibfnamefont {A.}~\bibnamefont {Gethin}},  \emph
		{et~al.},\ }\bibfield  {title} {\enquote {\bibinfo {title} {How unequal is
				europe? evidence from distributional national accounts, 1980-2017},}\
	}\href@noop {} {\bibfield  {journal} {\bibinfo  {journal} {WID. World working
				paper}\ }\textbf {\bibinfo {volume} {6}} (\bibinfo {year}
		{2019})}\BibitemShut {NoStop}%
	\bibitem [{\citenamefont {Chancel}\ \emph {et~al.}(2021)\citenamefont
		{Chancel}, \citenamefont {Piketty}, \citenamefont {Saez},\ and\ \citenamefont
		{Zucman}}]{chancel2021world}%
	\BibitemOpen
	\bibfield  {author} {\bibinfo {author} {\bibfnamefont {L.}~\bibnamefont
			{Chancel}}, \bibinfo {author} {\bibfnamefont {T.}~\bibnamefont {Piketty}},
		\bibinfo {author} {\bibfnamefont {E.}~\bibnamefont {Saez}}, \ and\ \bibinfo
		{author} {\bibfnamefont {G.}~\bibnamefont {Zucman}},\ }\bibfield  {title}
	{\enquote {\bibinfo {title} {World inequality report 2022},}\ }\href@noop {}
	{\bibfield  {journal} {\bibinfo  {journal} {World Inequality Lab}\ }
		(\bibinfo {year} {2021})}\BibitemShut {NoStop}%
	\bibitem [{\citenamefont {Sumner}(1906)}]{sumner:1906}%
	\BibitemOpen
	\bibfield  {author} {\bibinfo {author} {\bibfnamefont {W.~G.}\ \bibnamefont
			{Sumner}},\ }\bibfield  {title} {\enquote {\bibinfo {title} {Folways},}\
	}\href@noop {} {\bibfield  {journal} {\bibinfo  {journal} {Boston. Ginn}\ }
		(\bibinfo {year} {1906})}\BibitemShut {NoStop}%
	\bibitem [{\citenamefont {LeVine}\ and\ \citenamefont
		{Campbell}(1972)}]{levine:1972}%
	\BibitemOpen
	\bibfield  {author} {\bibinfo {author} {\bibfnamefont {R.~A.}\ \bibnamefont
			{LeVine}}\ and\ \bibinfo {author} {\bibfnamefont {D.~T.}\ \bibnamefont
			{Campbell}},\ }\bibfield  {title} {\enquote {\bibinfo {title}
			{Ethnocentrism},}\ }\href@noop {} {\bibfield  {journal} {\bibinfo  {journal}
			{New York: John Wiley}\ } (\bibinfo {year} {1972})}\BibitemShut {NoStop}%
	\bibitem [{\citenamefont {Hirschfeld}(1996)}]{Hirschfeld:1996}%
	\BibitemOpen
	\bibfield  {author} {\bibinfo {author} {\bibfnamefont {L.~A.}\ \bibnamefont
			{Hirschfeld}},\ }\bibfield  {title} {\enquote {\bibinfo {title} {Race in the
				making},}\ }\href@noop {} {\bibfield  {journal} {\bibinfo  {journal}
			{Cambridge, MA: MIT Press}\ } (\bibinfo {year} {1996})}\BibitemShut {NoStop}%
	\bibitem [{\citenamefont {Bernhard}, \citenamefont {Fischbacher},\ and\
		\citenamefont {Fehr}(2006)}]{bernhard:2006}%
	\BibitemOpen
	\bibfield  {author} {\bibinfo {author} {\bibfnamefont {H.}~\bibnamefont
			{Bernhard}}, \bibinfo {author} {\bibfnamefont {U.}~\bibnamefont
			{Fischbacher}}, \ and\ \bibinfo {author} {\bibfnamefont {E.}~\bibnamefont
			{Fehr}},\ }\bibfield  {title} {\enquote {\bibinfo {title} {Parochial altruism
				in humans},}\ }\href@noop {} {\bibfield  {journal} {\bibinfo  {journal}
			{Nature}\ }\textbf {\bibinfo {volume} {442}},\ \bibinfo {pages} {912--915}
		(\bibinfo {year} {2006})}\BibitemShut {NoStop}%
	\bibitem [{\citenamefont {Struch}\ and\ \citenamefont
		{Schwartz}(1989)}]{struch:1989}%
	\BibitemOpen
	\bibfield  {author} {\bibinfo {author} {\bibfnamefont {N.}~\bibnamefont
			{Struch}}\ and\ \bibinfo {author} {\bibfnamefont {S.~H.}\ \bibnamefont
			{Schwartz}},\ }\bibfield  {title} {\enquote {\bibinfo {title} {Intergroup
				aggression: Its predictors and distinctness from in-group bias},}\
	}\href@noop {} {\bibfield  {journal} {\bibinfo  {journal} {Journal of
				Personality and Social Psychology}\ }\textbf {\bibinfo {volume} {56}},\
		\bibinfo {pages} {346--373} (\bibinfo {year} {1989})}\BibitemShut {NoStop}%
	\bibitem [{\citenamefont {Allport}(1954)}]{allport1954nature}%
	\BibitemOpen
	\bibfield  {author} {\bibinfo {author} {\bibfnamefont {G.}~\bibnamefont
			{Allport}},\ }\href@noop {} {\emph {\bibinfo {title} {The nature of
				prejudice}}}\ (\bibinfo  {publisher} {Addison-Wesley},\ \bibinfo {year}
	{1954})\BibitemShut {NoStop}%
	\bibitem [{\citenamefont {Fiske}(2000)}]{fiske2000stereotyping}%
	\BibitemOpen
	\bibfield  {author} {\bibinfo {author} {\bibfnamefont {S.~T.}\ \bibnamefont
			{Fiske}},\ }\bibfield  {title} {\enquote {\bibinfo {title} {Stereotyping,
				prejudice, and discrimination at the seam between the centuries: Evolution,
				culture, mind, and brain},}\ }\href@noop {} {\bibfield  {journal} {\bibinfo
			{journal} {European Journal of Social Psychology}\ }\textbf {\bibinfo
			{volume} {30}},\ \bibinfo {pages} {299--322} (\bibinfo {year}
		{2000})}\BibitemShut {NoStop}%
	\bibitem [{\citenamefont {Fiske}(2002)}]{fiske2002we}%
	\BibitemOpen
	\bibfield  {author} {\bibinfo {author} {\bibfnamefont {S.~T.}\ \bibnamefont
			{Fiske}},\ }\bibfield  {title} {\enquote {\bibinfo {title} {What we know now
				about bias and intergroup conflict, the problem of the century},}\
	}\href@noop {} {\bibfield  {journal} {\bibinfo  {journal} {Current Directions
				in Psychological Science}\ }\textbf {\bibinfo {volume} {11}},\ \bibinfo
		{pages} {123--128} (\bibinfo {year} {2002})}\BibitemShut {NoStop}%
	\bibitem [{\citenamefont {Yamagishi}, \citenamefont {Jin},\ and\ \citenamefont
		{Kiyonari}(1999)}]{yamagishi1999bounded}%
	\BibitemOpen
	\bibfield  {author} {\bibinfo {author} {\bibfnamefont {T.}~\bibnamefont
			{Yamagishi}}, \bibinfo {author} {\bibfnamefont {N.}~\bibnamefont {Jin}}, \
		and\ \bibinfo {author} {\bibfnamefont {T.}~\bibnamefont {Kiyonari}},\
	}\bibfield  {title} {\enquote {\bibinfo {title} {Bounded generalized
				reciprocity: Ingroup boasting and ingroup favoritism},}\ }\href@noop {}
	{\bibfield  {journal} {\bibinfo  {journal} {Advances in group processes}\
		}\textbf {\bibinfo {volume} {16}},\ \bibinfo {pages} {161--197} (\bibinfo
		{year} {1999})}\BibitemShut {NoStop}%
	\bibitem [{\citenamefont {Fu}\ \emph {et~al.}(2012{\natexlab{a}})\citenamefont
		{Fu}, \citenamefont {Tarnita}, \citenamefont {Christakis}, \citenamefont
		{Wang}, \citenamefont {Rand},\ and\ \citenamefont {Nowak}}]{fu2012evolution}%
	\BibitemOpen
	\bibfield  {author} {\bibinfo {author} {\bibfnamefont {F.}~\bibnamefont
			{Fu}}, \bibinfo {author} {\bibfnamefont {C.~E.}\ \bibnamefont {Tarnita}},
		\bibinfo {author} {\bibfnamefont {N.~A.}\ \bibnamefont {Christakis}},
		\bibinfo {author} {\bibfnamefont {L.}~\bibnamefont {Wang}}, \bibinfo {author}
		{\bibfnamefont {D.~G.}\ \bibnamefont {Rand}}, \ and\ \bibinfo {author}
		{\bibfnamefont {M.~A.}\ \bibnamefont {Nowak}},\ }\bibfield  {title} {\enquote
		{\bibinfo {title} {Evolution of in-group favoritism},}\ }\href@noop {}
	{\bibfield  {journal} {\bibinfo  {journal} {Scientific reports}\ }\textbf
		{\bibinfo {volume} {2}},\ \bibinfo {pages} {1--6} (\bibinfo {year}
		{2012}{\natexlab{a}})}\BibitemShut {NoStop}%
	\bibitem [{\citenamefont {Balliet}, \citenamefont {Wu},\ and\ \citenamefont
		{De~Dreu}(2014)}]{balliet2014ingroup}%
	\BibitemOpen
	\bibfield  {author} {\bibinfo {author} {\bibfnamefont {D.}~\bibnamefont
			{Balliet}}, \bibinfo {author} {\bibfnamefont {J.}~\bibnamefont {Wu}}, \ and\
		\bibinfo {author} {\bibfnamefont {C.~K.}\ \bibnamefont {De~Dreu}},\
	}\bibfield  {title} {\enquote {\bibinfo {title} {Ingroup favoritism in
				cooperation: a meta-analysis.}}\ }\href@noop {} {\bibfield  {journal}
		{\bibinfo  {journal} {Psychological bulletin}\ }\textbf {\bibinfo {volume}
			{140}},\ \bibinfo {pages} {1556} (\bibinfo {year} {2014})}\BibitemShut
	{NoStop}%
	\bibitem [{\citenamefont {Ray}\ and\ \citenamefont {Lovejoy}(1986)}]{ray:1986}%
	\BibitemOpen
	\bibfield  {author} {\bibinfo {author} {\bibfnamefont {J.~J.}\ \bibnamefont
			{Ray}}\ and\ \bibinfo {author} {\bibfnamefont {F.~H.}\ \bibnamefont
			{Lovejoy}},\ }\bibfield  {title} {\enquote {\bibinfo {title} {The generality
				of racial prejudice},}\ }\href@noop {} {\bibfield  {journal} {\bibinfo
			{journal} {Journal of Social Psychology}\ }\textbf {\bibinfo {volume}
			{126}},\ \bibinfo {pages} {563--564} (\bibinfo {year} {1986})}\BibitemShut
	{NoStop}%
	\bibitem [{\citenamefont {Palmore}(1999)}]{palmore1999ageism}%
	\BibitemOpen
	\bibfield  {author} {\bibinfo {author} {\bibfnamefont {E.}~\bibnamefont
			{Palmore}},\ }\href@noop {} {\emph {\bibinfo {title} {Ageism: Negative and
				positive}}}\ (\bibinfo  {publisher} {Springer Publishing Company},\ \bibinfo
	{year} {1999})\BibitemShut {NoStop}%
	\bibitem [{\citenamefont {Cashdan}(2001)}]{cashdan:2001}%
	\BibitemOpen
	\bibfield  {author} {\bibinfo {author} {\bibfnamefont {E.}~\bibnamefont
			{Cashdan}},\ }\bibfield  {title} {\enquote {\bibinfo {title} {Ethnocentrism
				and xenophobia: A cross-cultural study},}\ }\href@noop {} {\bibfield
		{journal} {\bibinfo  {journal} {Current Anthropology}\ }\textbf {\bibinfo
			{volume} {42}},\ \bibinfo {pages} {760--765} (\bibinfo {year}
		{2001})}\BibitemShut {NoStop}%
	\bibitem [{\citenamefont {Bertrand}\ and\ \citenamefont
		{Duflo}(2017)}]{bertrand2017field}%
	\BibitemOpen
	\bibfield  {author} {\bibinfo {author} {\bibfnamefont {M.}~\bibnamefont
			{Bertrand}}\ and\ \bibinfo {author} {\bibfnamefont {E.}~\bibnamefont
			{Duflo}},\ }\bibfield  {title} {\enquote {\bibinfo {title} {Field experiments
				on discrimination},}\ }\href@noop {} {\bibfield  {journal} {\bibinfo
			{journal} {Handbook of economic field experiments}\ }\textbf {\bibinfo
			{volume} {1}},\ \bibinfo {pages} {309--393} (\bibinfo {year}
		{2017})}\BibitemShut {NoStop}%
	\bibitem [{\citenamefont {Ellemers}\ \emph {et~al.}(2018)\citenamefont
		{Ellemers} \emph {et~al.}}]{ellemers2018gender}%
	\BibitemOpen
	\bibfield  {author} {\bibinfo {author} {\bibfnamefont {N.}~\bibnamefont
			{Ellemers}} \emph {et~al.},\ }\bibfield  {title} {\enquote {\bibinfo {title}
			{Gender stereotypes},}\ }\href@noop {} {\bibfield  {journal} {\bibinfo
			{journal} {Annual review of psychology}\ }\textbf {\bibinfo {volume} {69}},\
		\bibinfo {pages} {275--298} (\bibinfo {year} {2018})}\BibitemShut {NoStop}%
	\bibitem [{\citenamefont {Dollard}\ \emph {et~al.}(1939)\citenamefont
		{Dollard}, \citenamefont {Miller}, \citenamefont {Doob}, \citenamefont
		{Mowrer},\ and\ \citenamefont {Sears}}]{dollard:1939}%
	\BibitemOpen
	\bibfield  {author} {\bibinfo {author} {\bibfnamefont {J.}~\bibnamefont
			{Dollard}}, \bibinfo {author} {\bibfnamefont {N.~E.}\ \bibnamefont {Miller}},
		\bibinfo {author} {\bibfnamefont {L.~W.}\ \bibnamefont {Doob}}, \bibinfo
		{author} {\bibfnamefont {O.~H.}\ \bibnamefont {Mowrer}}, \ and\ \bibinfo
		{author} {\bibfnamefont {R.~R.}\ \bibnamefont {Sears}},\ }\href@noop {}
	{\emph {\bibinfo {title} {Frustration and Aggression}}}\ (\bibinfo
	{publisher} {Yale University Press},\ \bibinfo {address} {New Haven,
		Connecticut},\ \bibinfo {year} {1939})\BibitemShut {NoStop}%
	\bibitem [{\citenamefont {Hovland}\ and\ \citenamefont
		{Sears}(1940)}]{hovland:1940}%
	\BibitemOpen
	\bibfield  {author} {\bibinfo {author} {\bibfnamefont {C.~I.}\ \bibnamefont
			{Hovland}}\ and\ \bibinfo {author} {\bibfnamefont {R.~S.}\ \bibnamefont
			{Sears}},\ }\bibfield  {title} {\enquote {\bibinfo {title} {Minor studies of
				aggression vi: Correlation of lynchings with economic indices},}\ }\href@noop
	{} {\bibfield  {journal} {\bibinfo  {journal} {J. Psych.}\ }\textbf {\bibinfo
			{volume} {9}},\ \bibinfo {pages} {301--310} (\bibinfo {year}
		{1940})}\BibitemShut {NoStop}%
	\bibitem [{\citenamefont {Lipset}(1959)}]{lipset1959democracy}%
	\BibitemOpen
	\bibfield  {author} {\bibinfo {author} {\bibfnamefont {S.~M.}\ \bibnamefont
			{Lipset}},\ }\bibfield  {title} {\enquote {\bibinfo {title} {Democracy and
				working-class authoritarianism},}\ }\href@noop {} {\bibfield  {journal}
		{\bibinfo  {journal} {American Sociological Review}\ ,\ \bibinfo {pages}
			{482--501}} (\bibinfo {year} {1959})}\BibitemShut {NoStop}%
	\bibitem [{\citenamefont {Levin}\ and\ \citenamefont
		{Levin}(1982)}]{levin:1982}%
	\BibitemOpen
	\bibfield  {author} {\bibinfo {author} {\bibfnamefont {J.}~\bibnamefont
			{Levin}}\ and\ \bibinfo {author} {\bibfnamefont {W.~C.}\ \bibnamefont
			{Levin}},\ }\href@noop {} {\emph {\bibinfo {title} {The Functions of
				Discrimination and Prejudice}}}\ (\bibinfo  {publisher} {Harper and Row},\
	\bibinfo {address} {New York},\ \bibinfo {year} {1982})\BibitemShut {NoStop}%
	\bibitem [{\citenamefont {Persell}, \citenamefont {Green},\ and\ \citenamefont
		{Gurevich}(2001)}]{persell2001civil}%
	\BibitemOpen
	\bibfield  {author} {\bibinfo {author} {\bibfnamefont {C.~H.}\ \bibnamefont
			{Persell}}, \bibinfo {author} {\bibfnamefont {A.}~\bibnamefont {Green}}, \
		and\ \bibinfo {author} {\bibfnamefont {L.}~\bibnamefont {Gurevich}},\
	}\bibfield  {title} {\enquote {\bibinfo {title} {Civil society, economic
				distress, and social tolerance},}\ }in\ \href@noop {} {\emph {\bibinfo
			{booktitle} {Sociological forum}}},\ Vol.~\bibinfo {volume} {16}\ (\bibinfo
	{organization} {Springer},\ \bibinfo {year} {2001})\ pp.\ \bibinfo {pages}
	{203--230}\BibitemShut {NoStop}%
	\bibitem [{\citenamefont {Gibson}(2002)}]{gibson:2002}%
	\BibitemOpen
	\bibfield  {author} {\bibinfo {author} {\bibfnamefont {J.~L.}\ \bibnamefont
			{Gibson}},\ }\bibfield  {title} {\enquote {\bibinfo {title} {{Becoming
					Tolerant? Short-Term Changes in Russian Political Culture}},}\ }\href@noop {}
	{\bibfield  {journal} {\bibinfo  {journal} {Brit. J. Pol. Sci.}\ }\textbf
		{\bibinfo {volume} {32}},\ \bibinfo {pages} {309--333} (\bibinfo {year}
		{2002})}\BibitemShut {NoStop}%
	\bibitem [{\citenamefont {Svallfors}(2006)}]{svallfors2006moral}%
	\BibitemOpen
	\bibfield  {author} {\bibinfo {author} {\bibfnamefont {S.}~\bibnamefont
			{Svallfors}},\ }\href@noop {} {\emph {\bibinfo {title} {The moral economy of
				class: Class and attitudes in comparative perspective}}}\ (\bibinfo
	{publisher} {Stanford University Press},\ \bibinfo {year} {2006})\BibitemShut
	{NoStop}%
	\bibitem [{\citenamefont {Knack}\ and\ \citenamefont
		{Keefer}(1997)}]{knack1997does}%
	\BibitemOpen
	\bibfield  {author} {\bibinfo {author} {\bibfnamefont {S.}~\bibnamefont
			{Knack}}\ and\ \bibinfo {author} {\bibfnamefont {P.}~\bibnamefont {Keefer}},\
	}\bibfield  {title} {\enquote {\bibinfo {title} {Does social capital have an
				economic payoff? a cross-country investigation},}\ }\href@noop {} {\bibfield
		{journal} {\bibinfo  {journal} {The Quarterly journal of economics}\ }\textbf
		{\bibinfo {volume} {112}},\ \bibinfo {pages} {1251--1288} (\bibinfo {year}
		{1997})}\BibitemShut {NoStop}%
	\bibitem [{\citenamefont {Uslaner}(2002)}]{uslaner2002moral}%
	\BibitemOpen
	\bibfield  {author} {\bibinfo {author} {\bibfnamefont {E.~M.}\ \bibnamefont
			{Uslaner}},\ }\href@noop {} {\emph {\bibinfo {title} {The Moral Foundations
				of Trust}}}\ (\bibinfo  {publisher} {Cambridge University Press},\ \bibinfo
	{year} {2002})\BibitemShut {NoStop}%
	\bibitem [{\citenamefont {Andersen}\ and\ \citenamefont
		{Fetner}(2008)}]{andersen2008economic}%
	\BibitemOpen
	\bibfield  {author} {\bibinfo {author} {\bibfnamefont {R.}~\bibnamefont
			{Andersen}}\ and\ \bibinfo {author} {\bibfnamefont {T.}~\bibnamefont
			{Fetner}},\ }\bibfield  {title} {\enquote {\bibinfo {title} {Economic
				inequality and intolerance: Attitudes toward homosexuality in 35
				democracies},}\ }\href@noop {} {\bibfield  {journal} {\bibinfo  {journal}
			{American Journal of Political Science}\ }\textbf {\bibinfo {volume} {52}},\
		\bibinfo {pages} {942--958} (\bibinfo {year} {2008})}\BibitemShut {NoStop}%
	\bibitem [{\citenamefont {Andersen}\ and\ \citenamefont
		{Curtis}(2012)}]{andersen2012polarizing}%
	\BibitemOpen
	\bibfield  {author} {\bibinfo {author} {\bibfnamefont {R.}~\bibnamefont
			{Andersen}}\ and\ \bibinfo {author} {\bibfnamefont {J.}~\bibnamefont
			{Curtis}},\ }\bibfield  {title} {\enquote {\bibinfo {title} {The polarizing
				effect of economic inequality on class identification: Evidence from 44
				countries},}\ }\href@noop {} {\bibfield  {journal} {\bibinfo  {journal}
			{Research in Social Stratification and Mobility}\ }\textbf {\bibinfo {volume}
			{30}},\ \bibinfo {pages} {129--141} (\bibinfo {year} {2012})}\BibitemShut
	{NoStop}%
	\bibitem [{\citenamefont {Nhim}, \citenamefont {Richter},\ and\ \citenamefont
		{Zhu}(2019)}]{nhim2019resilience}%
	\BibitemOpen
	\bibfield  {author} {\bibinfo {author} {\bibfnamefont {T.}~\bibnamefont
			{Nhim}}, \bibinfo {author} {\bibfnamefont {A.}~\bibnamefont {Richter}}, \
		and\ \bibinfo {author} {\bibfnamefont {X.}~\bibnamefont {Zhu}},\ }\bibfield
	{title} {\enquote {\bibinfo {title} {The resilience of social norms of
				cooperation under resource scarcity and inequality—an agent-based model on
				sharing water over two harvesting seasons},}\ }\href@noop {} {\bibfield
		{journal} {\bibinfo  {journal} {Ecological Complexity}\ }\textbf {\bibinfo
			{volume} {40}},\ \bibinfo {pages} {100709} (\bibinfo {year}
		{2019})}\BibitemShut {NoStop}%
	\bibitem [{\citenamefont {Masuda}(2012)}]{masuda:2012}%
	\BibitemOpen
	\bibfield  {author} {\bibinfo {author} {\bibfnamefont {N.}~\bibnamefont
			{Masuda}},\ }\bibfield  {title} {\enquote {\bibinfo {title} {Ingroup
				favoritism and intergroup cooperation under indirect reciprocity based on
				group reputation},}\ }\href@noop {} {\bibfield  {journal} {\bibinfo
			{journal} {J.\ Theor.\ Biol.}\ }\textbf {\bibinfo {volume} {311}},\ \bibinfo
		{pages} {8--18} (\bibinfo {year} {2012})}\BibitemShut {NoStop}%
	\bibitem [{\citenamefont {Nakamura}\ and\ \citenamefont
		{Masuda}(2012)}]{nakamura:2012}%
	\BibitemOpen
	\bibfield  {author} {\bibinfo {author} {\bibfnamefont {M.}~\bibnamefont
			{Nakamura}}\ and\ \bibinfo {author} {\bibfnamefont {N.}~\bibnamefont
			{Masuda}},\ }\bibfield  {title} {\enquote {\bibinfo {title} {Groupwise
				information sharing promotes ingroup favoritism in indirect reciprocity},}\
	}\href@noop {} {\bibfield  {journal} {\bibinfo  {journal} {BMC Evol. Biol.}\
		}\textbf {\bibinfo {volume} {12}},\ \bibinfo {pages} {213} (\bibinfo {year}
		{2012})}\BibitemShut {NoStop}%
	\bibitem [{\citenamefont {Oishi}, \citenamefont {Shimad},\ and\ \citenamefont
		{Ito}(2013)}]{oishi:2013}%
	\BibitemOpen
	\bibfield  {author} {\bibinfo {author} {\bibfnamefont {K.}~\bibnamefont
			{Oishi}}, \bibinfo {author} {\bibfnamefont {T.}~\bibnamefont {Shimad}}, \
		and\ \bibinfo {author} {\bibfnamefont {N.}~\bibnamefont {Ito}},\ }\bibfield
	{title} {\enquote {\bibinfo {title} {Group formation through indirect
				reciprocity},}\ }\href {\doibase 10.1103/PhysRevE.87.030801} {\bibfield
		{journal} {\bibinfo  {journal} {Phys. Rev. E}\ }\textbf {\bibinfo {volume}
			{87}},\ \bibinfo {pages} {030801} (\bibinfo {year} {2013})}\BibitemShut
	{NoStop}%
	\bibitem [{\citenamefont {Whitaker}, \citenamefont {Colombo},\ and\
		\citenamefont {Rand}(2018)}]{whitaker2018indirect}%
	\BibitemOpen
	\bibfield  {author} {\bibinfo {author} {\bibfnamefont {R.~M.}\ \bibnamefont
			{Whitaker}}, \bibinfo {author} {\bibfnamefont {G.~B.}\ \bibnamefont
			{Colombo}}, \ and\ \bibinfo {author} {\bibfnamefont {D.~G.}\ \bibnamefont
			{Rand}},\ }\bibfield  {title} {\enquote {\bibinfo {title} {Indirect
				reciprocity and the evolution of prejudicial groups},}\ }\href@noop {}
	{\bibfield  {journal} {\bibinfo  {journal} {Scientific reports}\ }\textbf
		{\bibinfo {volume} {8}},\ \bibinfo {pages} {1--14} (\bibinfo {year}
		{2018})}\BibitemShut {NoStop}%
	\bibitem [{\citenamefont {Martinez-Vaquero}\ and\ \citenamefont
		{Cuesta}(2014)}]{martinez:2014}%
	\BibitemOpen
	\bibfield  {author} {\bibinfo {author} {\bibfnamefont {L.~A.}\ \bibnamefont
			{Martinez-Vaquero}}\ and\ \bibinfo {author} {\bibfnamefont {J.~A.}\
			\bibnamefont {Cuesta}},\ }\bibfield  {title} {\enquote {\bibinfo {title}
			{Spreading of intolerance under economic stress: Results from a
				reputation-based model},}\ }\href@noop {} {\bibfield  {journal} {\bibinfo
			{journal} {Phys.\ Rev.\ E}\ }\textbf {\bibinfo {volume} {90}},\ \bibinfo
		{pages} {022805} (\bibinfo {year} {2014})}\BibitemShut {NoStop}%
	\bibitem [{\citenamefont {Sugden}(1986)}]{sugden:1986}%
	\BibitemOpen
	\bibfield  {author} {\bibinfo {author} {\bibfnamefont {R.}~\bibnamefont
			{Sugden}},\ }\href@noop {} {\emph {\bibinfo {title} {The economics of rights,
				co-operation and welfare}}}\ (\bibinfo  {publisher} {Basil Blackwell},\
	\bibinfo {address} {Oxford},\ \bibinfo {year} {1986})\BibitemShut {NoStop}%
	\bibitem [{\citenamefont {Alexander}(1987)}]{alexander:1987}%
	\BibitemOpen
	\bibfield  {author} {\bibinfo {author} {\bibfnamefont {R.~D.}\ \bibnamefont
			{Alexander}},\ }\href@noop {} {\emph {\bibinfo {title} {The Biology of Moral
				Systems}}}\ (\bibinfo  {publisher} {New York: Aldine de Gruyter},\ \bibinfo
	{year} {1987})\BibitemShut {NoStop}%
	\bibitem [{\citenamefont {Nowak}\ and\ \citenamefont
		{Sigmund}(1998)}]{nowak:1998}%
	\BibitemOpen
	\bibfield  {author} {\bibinfo {author} {\bibfnamefont {M.~A.}\ \bibnamefont
			{Nowak}}\ and\ \bibinfo {author} {\bibfnamefont {K.}~\bibnamefont
			{Sigmund}},\ }\bibfield  {title} {\enquote {\bibinfo {title} {Evolution of
				indirect reciprocity by image scoring},}\ }\href@noop {} {\bibfield
		{journal} {\bibinfo  {journal} {Nature}\ }\textbf {\bibinfo {volume} {393}},\
		\bibinfo {pages} {573--577} (\bibinfo {year} {1998})}\BibitemShut {NoStop}%
	\bibitem [{\citenamefont {Boyd}\ and\ \citenamefont
		{Richerson}(1989)}]{boyd1989evolution}%
	\BibitemOpen
	\bibfield  {author} {\bibinfo {author} {\bibfnamefont {R.}~\bibnamefont
			{Boyd}}\ and\ \bibinfo {author} {\bibfnamefont {P.~J.}\ \bibnamefont
			{Richerson}},\ }\bibfield  {title} {\enquote {\bibinfo {title} {The evolution
				of indirect reciprocity},}\ }\href@noop {} {\bibfield  {journal} {\bibinfo
			{journal} {Social networks}\ }\textbf {\bibinfo {volume} {11}},\ \bibinfo
		{pages} {213--236} (\bibinfo {year} {1989})}\BibitemShut {NoStop}%
	\bibitem [{\citenamefont {Leimar}\ and\ \citenamefont
		{Hammerstein}(2001)}]{leimar:2001}%
	\BibitemOpen
	\bibfield  {author} {\bibinfo {author} {\bibfnamefont {O.}~\bibnamefont
			{Leimar}}\ and\ \bibinfo {author} {\bibfnamefont {P.}~\bibnamefont
			{Hammerstein}},\ }\bibfield  {title} {\enquote {\bibinfo {title} {Evolution
				of cooperation through indirect reciprocity},}\ }\href@noop {} {\bibfield
		{journal} {\bibinfo  {journal} {Proc. R. Soc. Lond. B}\ }\textbf {\bibinfo
			{volume} {268}},\ \bibinfo {pages} {745--753} (\bibinfo {year}
		{2001})}\BibitemShut {NoStop}%
	\bibitem [{\citenamefont {Panchanathan}\ and\ \citenamefont
		{Boyd}(2003)}]{panchanathan:2003}%
	\BibitemOpen
	\bibfield  {author} {\bibinfo {author} {\bibfnamefont {K.}~\bibnamefont
			{Panchanathan}}\ and\ \bibinfo {author} {\bibfnamefont {R.}~\bibnamefont
			{Boyd}},\ }\bibfield  {title} {\enquote {\bibinfo {title} {A tale of two
				defectors: the importance of standing for evolution of indirect
				reciprocity},}\ }\href@noop {} {\bibfield  {journal} {\bibinfo  {journal}
			{J.\ Theor.\ Biol.}\ }\textbf {\bibinfo {volume} {224}},\ \bibinfo {pages}
		{115--126} (\bibinfo {year} {2003})}\BibitemShut {NoStop}%
	\bibitem [{\citenamefont {Nowak}\ and\ \citenamefont
		{Sigmund}(2005)}]{nowak:2005}%
	\BibitemOpen
	\bibfield  {author} {\bibinfo {author} {\bibfnamefont {M.~A.}\ \bibnamefont
			{Nowak}}\ and\ \bibinfo {author} {\bibfnamefont {K.}~\bibnamefont
			{Sigmund}},\ }\bibfield  {title} {\enquote {\bibinfo {title} {Evolution of
				indirect reciprocity},}\ }\href@noop {} {\bibfield  {journal} {\bibinfo
			{journal} {Nature}\ }\textbf {\bibinfo {volume} {437}},\ \bibinfo {pages}
		{1291--1298} (\bibinfo {year} {2005})}\BibitemShut {NoStop}%
	\bibitem [{\citenamefont {Nowak}(2006)}]{nowak:2006b}%
	\BibitemOpen
	\bibfield  {author} {\bibinfo {author} {\bibfnamefont {M.~A.}\ \bibnamefont
			{Nowak}},\ }\bibfield  {title} {\enquote {\bibinfo {title} {Five rules for
				the evolution of cooperation},}\ }\href@noop {} {\bibfield  {journal}
		{\bibinfo  {journal} {Science}\ }\textbf {\bibinfo {volume} {314}},\ \bibinfo
		{pages} {1560--1563} (\bibinfo {year} {2006})}\BibitemShut {NoStop}%
	\bibitem [{\citenamefont {Dufwenberg}\ \emph {et~al.}(2001)\citenamefont
		{Dufwenberg}, \citenamefont {Gneezy}, \citenamefont {G{\"u}th},\ and\
		\citenamefont {van Demme}}]{dufwenberg:2001}%
	\BibitemOpen
	\bibfield  {author} {\bibinfo {author} {\bibfnamefont {M.}~\bibnamefont
			{Dufwenberg}}, \bibinfo {author} {\bibfnamefont {U.}~\bibnamefont {Gneezy}},
		\bibinfo {author} {\bibfnamefont {W.}~\bibnamefont {G{\"u}th}}, \ and\
		\bibinfo {author} {\bibfnamefont {E.}~\bibnamefont {van Demme}},\ }\bibfield
	{title} {\enquote {\bibinfo {title} {Direct vs indirect reciprocity: An
				experiment},}\ }\href@noop {} {\bibfield  {journal} {\bibinfo  {journal}
			{Homo Oecon}\ }\textbf {\bibinfo {volume} {18}},\ \bibinfo {pages} {19--30}
		(\bibinfo {year} {2001})}\BibitemShut {NoStop}%
	\bibitem [{\citenamefont {Milinski}, \citenamefont {Semmann},\ and\
		\citenamefont {Krambeck}(2002)}]{milinski:2002}%
	\BibitemOpen
	\bibfield  {author} {\bibinfo {author} {\bibfnamefont {M.}~\bibnamefont
			{Milinski}}, \bibinfo {author} {\bibfnamefont {D.}~\bibnamefont {Semmann}}, \
		and\ \bibinfo {author} {\bibfnamefont {H.-J.}\ \bibnamefont {Krambeck}},\
	}\bibfield  {title} {\enquote {\bibinfo {title} {Reputation helps solve the
				tragedy of the commons},}\ }\href@noop {} {\bibfield  {journal} {\bibinfo
			{journal} {Nature}\ }\textbf {\bibinfo {volume} {415}},\ \bibinfo {pages}
		{424--426} (\bibinfo {year} {2002})}\BibitemShut {NoStop}%
	\bibitem [{\citenamefont {Panchanathan}\ and\ \citenamefont
		{Boyd}(2004)}]{panchanathan:2004}%
	\BibitemOpen
	\bibfield  {author} {\bibinfo {author} {\bibfnamefont {K.}~\bibnamefont
			{Panchanathan}}\ and\ \bibinfo {author} {\bibfnamefont {R.}~\bibnamefont
			{Boyd}},\ }\bibfield  {title} {\enquote {\bibinfo {title} {Indirect
				reciprocity can stabilise cooperation without the second-order free-rider
				problem},}\ }\href@noop {} {\bibfield  {journal} {\bibinfo  {journal}
			{Nature}\ }\textbf {\bibinfo {volume} {432}},\ \bibinfo {pages} {499--205}
		(\bibinfo {year} {2004})}\BibitemShut {NoStop}%
	\bibitem [{\citenamefont {Semmann}, \citenamefont {Krambeck},\ and\
		\citenamefont {Milinski}(2004)}]{semmann:2004}%
	\BibitemOpen
	\bibfield  {author} {\bibinfo {author} {\bibfnamefont {D.}~\bibnamefont
			{Semmann}}, \bibinfo {author} {\bibfnamefont {H.-J.}\ \bibnamefont
			{Krambeck}}, \ and\ \bibinfo {author} {\bibfnamefont {M.}~\bibnamefont
			{Milinski}},\ }\bibfield  {title} {\enquote {\bibinfo {title} {Strategic
				investment in reputation},}\ }\href@noop {} {\bibfield  {journal} {\bibinfo
			{journal} {J. Behav. Ecol. Sociobiol.}\ }\textbf {\bibinfo {volume} {56}},\
		\bibinfo {pages} {248--252} (\bibinfo {year} {2004})}\BibitemShut {NoStop}%
	\bibitem [{\citenamefont {Suzuki}\ and\ \citenamefont
		{Akiyama}(2007)}]{suzuki:2007}%
	\BibitemOpen
	\bibfield  {author} {\bibinfo {author} {\bibfnamefont {S.}~\bibnamefont
			{Suzuki}}\ and\ \bibinfo {author} {\bibfnamefont {E.}~\bibnamefont
			{Akiyama}},\ }\bibfield  {title} {\enquote {\bibinfo {title} {Evolution of
				indirect reciprocity in groups of various sizes and comparison with direct
				reciprocity},}\ }\href@noop {} {\bibfield  {journal} {\bibinfo  {journal}
			{J.\ Theor.\ Biol.}\ }\textbf {\bibinfo {volume} {245}},\ \bibinfo {pages}
		{539--552} (\bibinfo {year} {2007})}\BibitemShut {NoStop}%
	\bibitem [{\citenamefont {Yoeli}\ \emph {et~al.}(2013)\citenamefont {Yoeli},
		\citenamefont {Hoffman}, \citenamefont {Rand},\ and\ \citenamefont
		{Nowak}}]{yoeli2013powering}%
	\BibitemOpen
	\bibfield  {author} {\bibinfo {author} {\bibfnamefont {E.}~\bibnamefont
			{Yoeli}}, \bibinfo {author} {\bibfnamefont {M.}~\bibnamefont {Hoffman}},
		\bibinfo {author} {\bibfnamefont {D.~G.}\ \bibnamefont {Rand}}, \ and\
		\bibinfo {author} {\bibfnamefont {M.~A.}\ \bibnamefont {Nowak}},\ }\bibfield
	{title} {\enquote {\bibinfo {title} {Powering up with indirect reciprocity in
				a large-scale field experiment},}\ }\href@noop {} {\bibfield  {journal}
		{\bibinfo  {journal} {Proceedings of the National Academy of Sciences}\
		}\textbf {\bibinfo {volume} {110}},\ \bibinfo {pages} {10424--10429}
		(\bibinfo {year} {2013})}\BibitemShut {NoStop}%
	\bibitem [{\citenamefont {Trivers}(1971)}]{trivers:1971}%
	\BibitemOpen
	\bibfield  {author} {\bibinfo {author} {\bibfnamefont {R.~L.}\ \bibnamefont
			{Trivers}},\ }\bibfield  {title} {\enquote {\bibinfo {title} {{The evolution
					of reciprocal altruism}},}\ }\href@noop {} {\bibfield  {journal} {\bibinfo
			{journal} {Q. Rev. Biol.}\ }\textbf {\bibinfo {volume} {46}},\ \bibinfo
		{pages} {35--57} (\bibinfo {year} {1971})}\BibitemShut {NoStop}%
	\bibitem [{\citenamefont {Nowak}\ and\ \citenamefont
		{Roch}(2007)}]{nowak:2007}%
	\BibitemOpen
	\bibfield  {author} {\bibinfo {author} {\bibfnamefont {M.~A.}\ \bibnamefont
			{Nowak}}\ and\ \bibinfo {author} {\bibfnamefont {S.}~\bibnamefont {Roch}},\
	}\bibfield  {title} {\enquote {\bibinfo {title} {Upstream reciprocity and the
				evolution of gratitude},}\ }\href@noop {} {\bibfield  {journal} {\bibinfo
			{journal} {Proc. R. Soc. B}\ }\textbf {\bibinfo {volume} {274}},\ \bibinfo
		{pages} {605--610} (\bibinfo {year} {2007})}\BibitemShut {NoStop}%
	\bibitem [{\citenamefont {Ohtsuki}\ and\ \citenamefont
		{Iwasa}(2004)}]{ohtsuki:2004}%
	\BibitemOpen
	\bibfield  {author} {\bibinfo {author} {\bibfnamefont {H.}~\bibnamefont
			{Ohtsuki}}\ and\ \bibinfo {author} {\bibfnamefont {Y.}~\bibnamefont
			{Iwasa}},\ }\bibfield  {title} {\enquote {\bibinfo {title} {How should we
				define goodness?---reputation dynamics in indirect reciprocity},}\
	}\href@noop {} {\bibfield  {journal} {\bibinfo  {journal} {J.\ Theor.\
				Biol.}\ }\textbf {\bibinfo {volume} {231}},\ \bibinfo {pages} {107--120}
		(\bibinfo {year} {2004})}\BibitemShut {NoStop}%
	\bibitem [{\citenamefont {Brandt}\ and\ \citenamefont
		{Sigmund}(2004)}]{brandt:2004}%
	\BibitemOpen
	\bibfield  {author} {\bibinfo {author} {\bibfnamefont {H.}~\bibnamefont
			{Brandt}}\ and\ \bibinfo {author} {\bibfnamefont {K.}~\bibnamefont
			{Sigmund}},\ }\bibfield  {title} {\enquote {\bibinfo {title} {The logic of
				reprobation: assessment and action rules for indirect reciprocation},}\
	}\href@noop {} {\bibfield  {journal} {\bibinfo  {journal} {J.\ Theor.\
				Biol.}\ }\textbf {\bibinfo {volume} {231}},\ \bibinfo {pages} {475--486}
		(\bibinfo {year} {2004})}\BibitemShut {NoStop}%
	\bibitem [{\citenamefont {Fehr}(2004)}]{fehr2004don}%
	\BibitemOpen
	\bibfield  {author} {\bibinfo {author} {\bibfnamefont {E.}~\bibnamefont
			{Fehr}},\ }\bibfield  {title} {\enquote {\bibinfo {title} {Don't lose your
				reputation},}\ }\href@noop {} {\bibfield  {journal} {\bibinfo  {journal}
			{Nature}\ }\textbf {\bibinfo {volume} {432}},\ \bibinfo {pages} {449--450}
		(\bibinfo {year} {2004})}\BibitemShut {NoStop}%
	\bibitem [{\citenamefont {Ohtsuki}\ and\ \citenamefont
		{Iwasa}(2006)}]{ohtsuki:2006b}%
	\BibitemOpen
	\bibfield  {author} {\bibinfo {author} {\bibfnamefont {H.}~\bibnamefont
			{Ohtsuki}}\ and\ \bibinfo {author} {\bibfnamefont {Y.}~\bibnamefont
			{Iwasa}},\ }\bibfield  {title} {\enquote {\bibinfo {title} {The leading
				eight: Social norms that can maintain cooperation by indirect reciprocity},}\
	}\href@noop {} {\bibfield  {journal} {\bibinfo  {journal} {J.\ Theor.\
				Biol.}\ }\textbf {\bibinfo {volume} {239}},\ \bibinfo {pages} {435--444}
		(\bibinfo {year} {2006})}\BibitemShut {NoStop}%
	\bibitem [{\citenamefont {Martinez-Vaquero}\ and\ \citenamefont
		{Cuesta}(2013)}]{martinez:2013}%
	\BibitemOpen
	\bibfield  {author} {\bibinfo {author} {\bibfnamefont {L.~A.}\ \bibnamefont
			{Martinez-Vaquero}}\ and\ \bibinfo {author} {\bibfnamefont {J.~A.}\
			\bibnamefont {Cuesta}},\ }\bibfield  {title} {\enquote {\bibinfo {title}
			{Evolutionary stability and resistance to cheating in an indirect reciprocity
				model based on reputation},}\ }\href@noop {} {\bibfield  {journal} {\bibinfo
			{journal} {Phys.\ Rev.\ E}\ }\textbf {\bibinfo {volume} {87}},\ \bibinfo
		{pages} {052810} (\bibinfo {year} {2013})}\BibitemShut {NoStop}%
	\bibitem [{\citenamefont {Bolton}, \citenamefont {Katok},\ and\ \citenamefont
		{Ockenfels}(2004)}]{bolton2004effective}%
	\BibitemOpen
	\bibfield  {author} {\bibinfo {author} {\bibfnamefont {G.~E.}\ \bibnamefont
			{Bolton}}, \bibinfo {author} {\bibfnamefont {E.}~\bibnamefont {Katok}}, \
		and\ \bibinfo {author} {\bibfnamefont {A.}~\bibnamefont {Ockenfels}},\
	}\bibfield  {title} {\enquote {\bibinfo {title} {How effective are electronic
				reputation mechanisms? an experimental investigation},}\ }\href@noop {}
	{\bibfield  {journal} {\bibinfo  {journal} {Management science}\ }\textbf
		{\bibinfo {volume} {50}},\ \bibinfo {pages} {1587--1602} (\bibinfo {year}
		{2004})}\BibitemShut {NoStop}%
	\bibitem [{\citenamefont {J{\o}sang}, \citenamefont {Ismail},\ and\
		\citenamefont {Boyd}(2007)}]{josang2007survey}%
	\BibitemOpen
	\bibfield  {author} {\bibinfo {author} {\bibfnamefont {A.}~\bibnamefont
			{J{\o}sang}}, \bibinfo {author} {\bibfnamefont {R.}~\bibnamefont {Ismail}}, \
		and\ \bibinfo {author} {\bibfnamefont {C.}~\bibnamefont {Boyd}},\ }\bibfield
	{title} {\enquote {\bibinfo {title} {A survey of trust and reputation systems
				for online service provision},}\ }\href@noop {} {\bibfield  {journal}
		{\bibinfo  {journal} {Decision support systems}\ }\textbf {\bibinfo {volume}
			{43}},\ \bibinfo {pages} {618--644} (\bibinfo {year} {2007})}\BibitemShut
	{NoStop}%
	\bibitem [{\citenamefont {Fishman}(2003)}]{fishman:2003}%
	\BibitemOpen
	\bibfield  {author} {\bibinfo {author} {\bibfnamefont {M.~A.}\ \bibnamefont
			{Fishman}},\ }\bibfield  {title} {\enquote {\bibinfo {title} {Indirect
				reciprocity among imperfect individuals},}\ }\href@noop {} {\bibfield
		{journal} {\bibinfo  {journal} {J.\ Theor.\ Biol.}\ }\textbf {\bibinfo
			{volume} {225}},\ \bibinfo {pages} {285--292} (\bibinfo {year}
		{2003})}\BibitemShut {NoStop}%
	\bibitem [{\citenamefont {Lotem}, \citenamefont {Fishman},\ and\ \citenamefont
		{Stone}(1999)}]{lotem:1999}%
	\BibitemOpen
	\bibfield  {author} {\bibinfo {author} {\bibfnamefont {A.}~\bibnamefont
			{Lotem}}, \bibinfo {author} {\bibfnamefont {M.~A.}\ \bibnamefont {Fishman}},
		\ and\ \bibinfo {author} {\bibfnamefont {L.}~\bibnamefont {Stone}},\
	}\bibfield  {title} {\enquote {\bibinfo {title} {Evolution of cooperation
				between individuals},}\ }\href@noop {} {\bibfield  {journal} {\bibinfo
			{journal} {Nature}\ }\textbf {\bibinfo {volume} {400}},\ \bibinfo {pages}
		{226--227} (\bibinfo {year} {1999})}\BibitemShut {NoStop}%
	\bibitem [{\citenamefont {Gini}(1936)}]{gini1936measure}%
	\BibitemOpen
	\bibfield  {author} {\bibinfo {author} {\bibfnamefont {C.}~\bibnamefont
			{Gini}},\ }\bibfield  {title} {\enquote {\bibinfo {title} {On the measure of
				concentration with special reference to income and statistics},}\ }\href@noop
	{} {\bibfield  {journal} {\bibinfo  {journal} {Colorado College Publication,
				General Series}\ }\textbf {\bibinfo {volume} {208}},\ \bibinfo {pages}
		{73--79} (\bibinfo {year} {1936})}\BibitemShut {NoStop}%
	\bibitem [{\citenamefont {Smith}\ and\ \citenamefont
		{Price}(1973)}]{smith1973logic}%
	\BibitemOpen
	\bibfield  {author} {\bibinfo {author} {\bibfnamefont {J.}~\bibnamefont
			{Smith}}\ and\ \bibinfo {author} {\bibfnamefont {G.~R.}\ \bibnamefont
			{Price}},\ }\bibfield  {title} {\enquote {\bibinfo {title} {The logic of
				animal conflict},}\ }\href@noop {} {\bibfield  {journal} {\bibinfo  {journal}
			{Nature}\ }\textbf {\bibinfo {volume} {246}},\ \bibinfo {pages} {15--18}
		(\bibinfo {year} {1973})}\BibitemShut {NoStop}%
	\bibitem [{\citenamefont {Hackel}\ and\ \citenamefont
		{Zaki}(2018)}]{hackel2018propagation}%
	\BibitemOpen
	\bibfield  {author} {\bibinfo {author} {\bibfnamefont {L.~M.}\ \bibnamefont
			{Hackel}}\ and\ \bibinfo {author} {\bibfnamefont {J.}~\bibnamefont {Zaki}},\
	}\bibfield  {title} {\enquote {\bibinfo {title} {Propagation of economic
				inequality through reciprocity and reputation},}\ }\href@noop {} {\bibfield
		{journal} {\bibinfo  {journal} {Psychological science}\ }\textbf {\bibinfo
			{volume} {29}},\ \bibinfo {pages} {604--613} (\bibinfo {year}
		{2018})}\BibitemShut {NoStop}%
	\bibitem [{\citenamefont {Gangadharan}\ \emph {et~al.}(2019)\citenamefont
		{Gangadharan}, \citenamefont {Grossman}, \citenamefont {Molle},\ and\
		\citenamefont {Vecci}}]{gangadharan2019impact}%
	\BibitemOpen
	\bibfield  {author} {\bibinfo {author} {\bibfnamefont {L.}~\bibnamefont
			{Gangadharan}}, \bibinfo {author} {\bibfnamefont {P.~J.}\ \bibnamefont
			{Grossman}}, \bibinfo {author} {\bibfnamefont {M.~K.}\ \bibnamefont {Molle}},
		\ and\ \bibinfo {author} {\bibfnamefont {J.}~\bibnamefont {Vecci}},\
	}\bibfield  {title} {\enquote {\bibinfo {title} {Impact of social identity
				and inequality on antisocial behaviour},}\ }\href@noop {} {\bibfield
		{journal} {\bibinfo  {journal} {European Economic Review}\ }\textbf {\bibinfo
			{volume} {119}},\ \bibinfo {pages} {199--215} (\bibinfo {year}
		{2019})}\BibitemShut {NoStop}%
	\bibitem [{\citenamefont {Nowak}, \citenamefont {Sigmund},\ and\ \citenamefont
		{El-Sedy}(1995)}]{nowak:1995}%
	\BibitemOpen
	\bibfield  {author} {\bibinfo {author} {\bibfnamefont {M.~A.}\ \bibnamefont
			{Nowak}}, \bibinfo {author} {\bibfnamefont {K.}~\bibnamefont {Sigmund}}, \
		and\ \bibinfo {author} {\bibfnamefont {E.}~\bibnamefont {El-Sedy}},\
	}\bibfield  {title} {\enquote {\bibinfo {title} {Automata, repeated games and
				noise},}\ }\href@noop {} {\bibfield  {journal} {\bibinfo  {journal} {J. Math.
				Biol.}\ }\textbf {\bibinfo {volume} {33}},\ \bibinfo {pages} {703--722}
		(\bibinfo {year} {1995})}\BibitemShut {NoStop}%
	\bibitem [{\citenamefont {Martinez-Vaquero}, \citenamefont {Cuesta},\ and\
		\citenamefont {S\'anchez}(2012)}]{martinez:2012}%
	\BibitemOpen
	\bibfield  {author} {\bibinfo {author} {\bibfnamefont {L.~A.}\ \bibnamefont
			{Martinez-Vaquero}}, \bibinfo {author} {\bibfnamefont {J.~A.}\ \bibnamefont
			{Cuesta}}, \ and\ \bibinfo {author} {\bibfnamefont {A.}~\bibnamefont
			{S\'anchez}},\ }\bibfield  {title} {\enquote {\bibinfo {title} {Generosity
				pays in the presence of direct reciprocity: A comprehensive study of
				2$\times$2 repeated games},}\ }\href@noop {} {\bibfield  {journal} {\bibinfo
			{journal} {PLoS ONE}\ }\textbf {\bibinfo {volume} {7}},\ \bibinfo {pages}
		{e35135} (\bibinfo {year} {2012})}\BibitemShut {NoStop}%
	\bibitem [{\citenamefont {Mani}\ \emph {et~al.}(2013)\citenamefont {Mani},
		\citenamefont {Mullainathan}, \citenamefont {Shafir},\ and\ \citenamefont
		{Zhao}}]{mani2013poverty}%
	\BibitemOpen
	\bibfield  {author} {\bibinfo {author} {\bibfnamefont {A.}~\bibnamefont
			{Mani}}, \bibinfo {author} {\bibfnamefont {S.}~\bibnamefont {Mullainathan}},
		\bibinfo {author} {\bibfnamefont {E.}~\bibnamefont {Shafir}}, \ and\ \bibinfo
		{author} {\bibfnamefont {J.}~\bibnamefont {Zhao}},\ }\bibfield  {title}
	{\enquote {\bibinfo {title} {Poverty impedes cognitive function},}\
	}\href@noop {} {\bibfield  {journal} {\bibinfo  {journal} {science}\ }\textbf
		{\bibinfo {volume} {341}},\ \bibinfo {pages} {976--980} (\bibinfo {year}
		{2013})}\BibitemShut {NoStop}%
	\bibitem [{\citenamefont {Santos}, \citenamefont {Pacheco},\ and\ \citenamefont
		{Santos}(2021)}]{santos2021complexity}%
	\BibitemOpen
	\bibfield  {author} {\bibinfo {author} {\bibfnamefont {F.~P.}\ \bibnamefont
			{Santos}}, \bibinfo {author} {\bibfnamefont {J.~M.}\ \bibnamefont {Pacheco}},
		\ and\ \bibinfo {author} {\bibfnamefont {F.~C.}\ \bibnamefont {Santos}},\
	}\bibfield  {title} {\enquote {\bibinfo {title} {The complexity of human
				cooperation under indirect reciprocity},}\ }\href@noop {} {\bibfield
		{journal} {\bibinfo  {journal} {Philosophical Transactions of the Royal
				Society B}\ }\textbf {\bibinfo {volume} {376}},\ \bibinfo {pages} {20200291}
		(\bibinfo {year} {2021})}\BibitemShut {NoStop}%
	\bibitem [{\citenamefont {Fu}\ \emph {et~al.}(2012{\natexlab{b}})\citenamefont
		{Fu}, \citenamefont {Tarnita}, \citenamefont {Christakis}, \citenamefont
		{Wang}, \citenamefont {Rand},\ and\ \citenamefont {Nowak}}]{fu:2012}%
	\BibitemOpen
	\bibfield  {author} {\bibinfo {author} {\bibfnamefont {F.}~\bibnamefont
			{Fu}}, \bibinfo {author} {\bibfnamefont {C.~E.}\ \bibnamefont {Tarnita}},
		\bibinfo {author} {\bibfnamefont {N.~A.}\ \bibnamefont {Christakis}},
		\bibinfo {author} {\bibfnamefont {L.}~\bibnamefont {Wang}}, \bibinfo {author}
		{\bibfnamefont {D.~G.}\ \bibnamefont {Rand}}, \ and\ \bibinfo {author}
		{\bibfnamefont {M.~A.}\ \bibnamefont {Nowak}},\ }\bibfield  {title} {\enquote
		{\bibinfo {title} {Evolution of in-group favoritism},}\ }\href@noop {}
	{\bibfield  {journal} {\bibinfo  {journal} {Sci Rep.}\ }\textbf {\bibinfo
			{volume} {2}},\ \bibinfo {pages} {460} (\bibinfo {year}
		{2012}{\natexlab{b}})}\BibitemShut {NoStop}%
	\bibitem [{\citenamefont {Zhang}\ \emph {et~al.}(2021)\citenamefont {Zhang},
		\citenamefont {Huang}, \citenamefont {Li}, \citenamefont {Dai},\ and\
		\citenamefont {Yang}}]{zhang2021effects}%
	\BibitemOpen
	\bibfield  {author} {\bibinfo {author} {\bibfnamefont {L.}~\bibnamefont
			{Zhang}}, \bibinfo {author} {\bibfnamefont {C.}~\bibnamefont {Huang}},
		\bibinfo {author} {\bibfnamefont {H.}~\bibnamefont {Li}}, \bibinfo {author}
		{\bibfnamefont {Q.}~\bibnamefont {Dai}}, \ and\ \bibinfo {author}
		{\bibfnamefont {J.}~\bibnamefont {Yang}},\ }\bibfield  {title} {\enquote
		{\bibinfo {title} {Effects of directional migration for pursuit of profitable
				circumstances in evolutionary games},}\ }\href@noop {} {\bibfield  {journal}
		{\bibinfo  {journal} {Chaos, Solitons \& Fractals}\ }\textbf {\bibinfo
			{volume} {144}},\ \bibinfo {pages} {110709} (\bibinfo {year}
		{2021})}\BibitemShut {NoStop}%
	\bibitem [{\citenamefont {Oliveira}\ \emph {et~al.}(2022)\citenamefont
		{Oliveira}, \citenamefont {Karimi}, \citenamefont {Zens}, \citenamefont
		{Schaible}, \citenamefont {G{\'e}nois},\ and\ \citenamefont
		{Strohmaier}}]{oliveira2022group}%
	\BibitemOpen
	\bibfield  {author} {\bibinfo {author} {\bibfnamefont {M.}~\bibnamefont
			{Oliveira}}, \bibinfo {author} {\bibfnamefont {F.}~\bibnamefont {Karimi}},
		\bibinfo {author} {\bibfnamefont {M.}~\bibnamefont {Zens}}, \bibinfo {author}
		{\bibfnamefont {J.}~\bibnamefont {Schaible}}, \bibinfo {author}
		{\bibfnamefont {M.}~\bibnamefont {G{\'e}nois}}, \ and\ \bibinfo {author}
		{\bibfnamefont {M.}~\bibnamefont {Strohmaier}},\ }\bibfield  {title}
	{\enquote {\bibinfo {title} {Group mixing drives inequality in face-to-face
				gatherings},}\ }\href@noop {} {\bibfield  {journal} {\bibinfo  {journal}
			{Communications Physics}\ }\textbf {\bibinfo {volume} {5}},\ \bibinfo {pages}
		{1--9} (\bibinfo {year} {2022})}\BibitemShut {NoStop}%
	\bibitem [{\citenamefont {Murase}, \citenamefont {Kim},\ and\ \citenamefont
		{Baek}(2022)}]{murase2022social}%
	\BibitemOpen
	\bibfield  {author} {\bibinfo {author} {\bibfnamefont {Y.}~\bibnamefont
			{Murase}}, \bibinfo {author} {\bibfnamefont {M.}~\bibnamefont {Kim}}, \ and\
		\bibinfo {author} {\bibfnamefont {S.~K.}\ \bibnamefont {Baek}},\ }\bibfield
	{title} {\enquote {\bibinfo {title} {Social norms in indirect reciprocity
				with ternary reputations},}\ }\href@noop {} {\bibfield  {journal} {\bibinfo
			{journal} {Scientific Reports}\ }\textbf {\bibinfo {volume} {12}},\ \bibinfo
		{pages} {1--15} (\bibinfo {year} {2022})}\BibitemShut {NoStop}%
	\bibitem [{\citenamefont {Hilbe}\ \emph {et~al.}(2018)\citenamefont {Hilbe},
		\citenamefont {Schmid}, \citenamefont {Tkadlec}, \citenamefont {Chatterjee},\
		and\ \citenamefont {Nowak}}]{hilbe2018indirect}%
	\BibitemOpen
	\bibfield  {author} {\bibinfo {author} {\bibfnamefont {C.}~\bibnamefont
			{Hilbe}}, \bibinfo {author} {\bibfnamefont {L.}~\bibnamefont {Schmid}},
		\bibinfo {author} {\bibfnamefont {J.}~\bibnamefont {Tkadlec}}, \bibinfo
		{author} {\bibfnamefont {K.}~\bibnamefont {Chatterjee}}, \ and\ \bibinfo
		{author} {\bibfnamefont {M.~A.}\ \bibnamefont {Nowak}},\ }\bibfield  {title}
	{\enquote {\bibinfo {title} {Indirect reciprocity with private, noisy, and
				incomplete information},}\ }\href@noop {} {\bibfield  {journal} {\bibinfo
			{journal} {Proceedings of the national academy of sciences}\ }\textbf
		{\bibinfo {volume} {115}},\ \bibinfo {pages} {12241--12246} (\bibinfo {year}
		{2018})}\BibitemShut {NoStop}%
	\bibitem [{\citenamefont {Martinez-Vaquero}, \citenamefont {Santos},\ and\
		\citenamefont {Trianni}(2020)}]{martinez2020signalling}%
	\BibitemOpen
	\bibfield  {author} {\bibinfo {author} {\bibfnamefont {L.~A.}\ \bibnamefont
			{Martinez-Vaquero}}, \bibinfo {author} {\bibfnamefont {F.~C.}\ \bibnamefont
			{Santos}}, \ and\ \bibinfo {author} {\bibfnamefont {V.}~\bibnamefont
			{Trianni}},\ }\bibfield  {title} {\enquote {\bibinfo {title} {Signalling
				boosts the evolution of cooperation in repeated group interactions},}\
	}\href@noop {} {\bibfield  {journal} {\bibinfo  {journal} {Journal of the
				Royal Society Interface}\ }\textbf {\bibinfo {volume} {17}},\ \bibinfo
		{pages} {20200635} (\bibinfo {year} {2020})}\BibitemShut {NoStop}%
	\bibitem [{\citenamefont {Martinez-Vaquero}\ \emph {et~al.}(2015)\citenamefont
		{Martinez-Vaquero}, \citenamefont {Han}, \citenamefont {Pereira},\ and\
		\citenamefont {Lenaerts}}]{martinez2015apology}%
	\BibitemOpen
	\bibfield  {author} {\bibinfo {author} {\bibfnamefont {L.~A.}\ \bibnamefont
			{Martinez-Vaquero}}, \bibinfo {author} {\bibfnamefont {T.~A.}\ \bibnamefont
			{Han}}, \bibinfo {author} {\bibfnamefont {L.~M.}\ \bibnamefont {Pereira}}, \
		and\ \bibinfo {author} {\bibfnamefont {T.}~\bibnamefont {Lenaerts}},\
	}\bibfield  {title} {\enquote {\bibinfo {title} {Apology and forgiveness
				evolve to resolve failures in cooperative agreements},}\ }\href@noop {}
	{\bibfield  {journal} {\bibinfo  {journal} {Scientific reports}\ }\textbf
		{\bibinfo {volume} {5}},\ \bibinfo {pages} {1--12} (\bibinfo {year}
		{2015})}\BibitemShut {NoStop}%
	\bibitem [{\citenamefont {Karimi}\ \emph {et~al.}(2018)\citenamefont {Karimi},
		\citenamefont {G{\'e}nois}, \citenamefont {Wagner}, \citenamefont {Singer},\
		and\ \citenamefont {Strohmaier}}]{karimi2018homophily}%
	\BibitemOpen
	\bibfield  {author} {\bibinfo {author} {\bibfnamefont {F.}~\bibnamefont
			{Karimi}}, \bibinfo {author} {\bibfnamefont {M.}~\bibnamefont {G{\'e}nois}},
		\bibinfo {author} {\bibfnamefont {C.}~\bibnamefont {Wagner}}, \bibinfo
		{author} {\bibfnamefont {P.}~\bibnamefont {Singer}}, \ and\ \bibinfo {author}
		{\bibfnamefont {M.}~\bibnamefont {Strohmaier}},\ }\bibfield  {title}
	{\enquote {\bibinfo {title} {Homophily influences ranking of minorities in
				social networks},}\ }\href@noop {} {\bibfield  {journal} {\bibinfo  {journal}
			{Scientific reports}\ }\textbf {\bibinfo {volume} {8}},\ \bibinfo {pages}
		{1--12} (\bibinfo {year} {2018})}\BibitemShut {NoStop}%
	\bibitem [{\citenamefont {Hu}\ \emph {et~al.}(2021)\citenamefont {Hu},
		\citenamefont {Li}, \citenamefont {Wang}, \citenamefont {Xia}, \citenamefont
		{Wang},\ and\ \citenamefont {Perc}}]{hu2021adaptive}%
	\BibitemOpen
	\bibfield  {author} {\bibinfo {author} {\bibfnamefont {Z.}~\bibnamefont
			{Hu}}, \bibinfo {author} {\bibfnamefont {X.}~\bibnamefont {Li}}, \bibinfo
		{author} {\bibfnamefont {J.}~\bibnamefont {Wang}}, \bibinfo {author}
		{\bibfnamefont {C.}~\bibnamefont {Xia}}, \bibinfo {author} {\bibfnamefont
			{Z.}~\bibnamefont {Wang}}, \ and\ \bibinfo {author} {\bibfnamefont
			{M.}~\bibnamefont {Perc}},\ }\bibfield  {title} {\enquote {\bibinfo {title}
			{Adaptive reputation promotes trust in social networks},}\ }\href@noop {}
	{\bibfield  {journal} {\bibinfo  {journal} {IEEE Transactions on Network
				Science and Engineering}\ }\textbf {\bibinfo {volume} {8}},\ \bibinfo {pages}
		{3087--3098} (\bibinfo {year} {2021})}\BibitemShut {NoStop}%
	\bibitem [{\citenamefont {DiMaggio}\ and\ \citenamefont
		{Garip}(2012)}]{dimaggio2012network}%
	\BibitemOpen
	\bibfield  {author} {\bibinfo {author} {\bibfnamefont {P.}~\bibnamefont
			{DiMaggio}}\ and\ \bibinfo {author} {\bibfnamefont {F.}~\bibnamefont
			{Garip}},\ }\bibfield  {title} {\enquote {\bibinfo {title} {Network effects
				and social inequality},}\ }\href@noop {} {\bibfield  {journal} {\bibinfo
			{journal} {Annual review of sociology}\ } (\bibinfo {year}
		{2012})}\BibitemShut {NoStop}%
	\bibitem [{\citenamefont {Kas}, \citenamefont {Corten},\ and\ \citenamefont
		{van~de Rijt}(2022)}]{kas2022role}%
	\BibitemOpen
	\bibfield  {author} {\bibinfo {author} {\bibfnamefont {J.}~\bibnamefont
			{Kas}}, \bibinfo {author} {\bibfnamefont {R.}~\bibnamefont {Corten}}, \ and\
		\bibinfo {author} {\bibfnamefont {A.}~\bibnamefont {van~de Rijt}},\
	}\bibfield  {title} {\enquote {\bibinfo {title} {The role of reputation
				systems in digital discrimination},}\ }\href@noop {} {\bibfield  {journal}
		{\bibinfo  {journal} {Socio-economic review}\ }\textbf {\bibinfo {volume}
			{20}},\ \bibinfo {pages} {1905--1932} (\bibinfo {year} {2022})}\BibitemShut
	{NoStop}%
	\bibitem [{\citenamefont {Browman}\ \emph {et~al.}(2019)\citenamefont
		{Browman}, \citenamefont {Destin}, \citenamefont {Kearney},\ and\
		\citenamefont {Levine}}]{browman2019economic}%
	\BibitemOpen
	\bibfield  {author} {\bibinfo {author} {\bibfnamefont {A.~S.}\ \bibnamefont
			{Browman}}, \bibinfo {author} {\bibfnamefont {M.}~\bibnamefont {Destin}},
		\bibinfo {author} {\bibfnamefont {M.~S.}\ \bibnamefont {Kearney}}, \ and\
		\bibinfo {author} {\bibfnamefont {P.~B.}\ \bibnamefont {Levine}},\ }\bibfield
	{title} {\enquote {\bibinfo {title} {How economic inequality shapes mobility
				expectations and behaviour in disadvantaged youth},}\ }\href@noop {}
	{\bibfield  {journal} {\bibinfo  {journal} {Nature Human Behaviour}\ }\textbf
		{\bibinfo {volume} {3}},\ \bibinfo {pages} {214--220} (\bibinfo {year}
		{2019})}\BibitemShut {NoStop}%
	\bibitem [{\citenamefont {Jetten}\ \emph {et~al.}(2017)\citenamefont {Jetten},
		\citenamefont {Wang}, \citenamefont {Steffens}, \citenamefont {Mols},
		\citenamefont {Peters},\ and\ \citenamefont {Verkuyten}}]{jetten2017social}%
	\BibitemOpen
	\bibfield  {author} {\bibinfo {author} {\bibfnamefont {J.}~\bibnamefont
			{Jetten}}, \bibinfo {author} {\bibfnamefont {Z.}~\bibnamefont {Wang}},
		\bibinfo {author} {\bibfnamefont {N.~K.}\ \bibnamefont {Steffens}}, \bibinfo
		{author} {\bibfnamefont {F.}~\bibnamefont {Mols}}, \bibinfo {author}
		{\bibfnamefont {K.}~\bibnamefont {Peters}}, \ and\ \bibinfo {author}
		{\bibfnamefont {M.}~\bibnamefont {Verkuyten}},\ }\bibfield  {title} {\enquote
		{\bibinfo {title} {A social identity analysis of responses to economic
				inequality},}\ }\href@noop {} {\bibfield  {journal} {\bibinfo  {journal}
			{Current Opinion in Psychology}\ }\textbf {\bibinfo {volume} {18}},\ \bibinfo
		{pages} {1--5} (\bibinfo {year} {2017})}\BibitemShut {NoStop}%
	\bibitem [{\citenamefont {Jetten}\ and\ \citenamefont
		{Peters}(2019)}]{jetten2019social}%
	\BibitemOpen
	\bibfield  {author} {\bibinfo {author} {\bibfnamefont {J.}~\bibnamefont
			{Jetten}}\ and\ \bibinfo {author} {\bibfnamefont {K.}~\bibnamefont
			{Peters}},\ }\href@noop {} {\emph {\bibinfo {title} {The social psychology of
				inequality}}}\ (\bibinfo  {publisher} {Springer},\ \bibinfo {year}
	{2019})\BibitemShut {NoStop}%
	\bibitem [{\citenamefont {Bowles}\ \emph {et~al.}(2012)\citenamefont {Bowles},
		\citenamefont {Fong}, \citenamefont {Gintis},\ and\ \citenamefont
		{Pagano}}]{bowles2012new}%
	\BibitemOpen
	\bibfield  {author} {\bibinfo {author} {\bibfnamefont {S.}~\bibnamefont
			{Bowles}}, \bibinfo {author} {\bibfnamefont {C.~M.}\ \bibnamefont {Fong}},
		\bibinfo {author} {\bibfnamefont {H.}~\bibnamefont {Gintis}}, \ and\ \bibinfo
		{author} {\bibfnamefont {U.}~\bibnamefont {Pagano}},\ }\href@noop {} {\emph
		{\bibinfo {title} {The new economics of inequality and redistribution}}}\
	(\bibinfo  {publisher} {Cambridge University Press},\ \bibinfo {year}
	{2012})\BibitemShut {NoStop}%
	\bibitem [{\citenamefont {Russett}(1964)}]{russett1964inequality}%
	\BibitemOpen
	\bibfield  {author} {\bibinfo {author} {\bibfnamefont {B.~M.}\ \bibnamefont
			{Russett}},\ }\bibfield  {title} {\enquote {\bibinfo {title} {Inequality and
				instability: The relation of land tenure to politics},}\ }\href@noop {}
	{\bibfield  {journal} {\bibinfo  {journal} {World Politics}\ }\textbf
		{\bibinfo {volume} {16}},\ \bibinfo {pages} {442--454} (\bibinfo {year}
		{1964})}\BibitemShut {NoStop}%
	\bibitem [{\citenamefont {Huntington}(1968)}]{huntington1968political}%
	\BibitemOpen
	\bibfield  {author} {\bibinfo {author} {\bibfnamefont {S.~P.}\ \bibnamefont
			{Huntington}},\ }\href@noop {} {\emph {\bibinfo {title} {Political order in
				changing societies}}}\ (\bibinfo  {publisher} {Yale University Press},\
	\bibinfo {year} {1968})\BibitemShut {NoStop}%
	\bibitem [{\citenamefont {Alesina}\ and\ \citenamefont
		{Perotti}(1996)}]{alesina1996income}%
	\BibitemOpen
	\bibfield  {author} {\bibinfo {author} {\bibfnamefont {A.}~\bibnamefont
			{Alesina}}\ and\ \bibinfo {author} {\bibfnamefont {R.}~\bibnamefont
			{Perotti}},\ }\bibfield  {title} {\enquote {\bibinfo {title} {Income
				distribution, political instability, and investment},}\ }\href@noop {}
	{\bibfield  {journal} {\bibinfo  {journal} {European economic review}\
		}\textbf {\bibinfo {volume} {40}},\ \bibinfo {pages} {1203--1228} (\bibinfo
		{year} {1996})}\BibitemShut {NoStop}%
	\bibitem [{\citenamefont {Perotti}(1996)}]{perotti1996growth}%
	\BibitemOpen
	\bibfield  {author} {\bibinfo {author} {\bibfnamefont {R.}~\bibnamefont
			{Perotti}},\ }\bibfield  {title} {\enquote {\bibinfo {title} {Growth, income
				distribution, and democracy: What the data say},}\ }\href@noop {} {\bibfield
		{journal} {\bibinfo  {journal} {Journal of Economic growth}\ }\textbf
		{\bibinfo {volume} {1}},\ \bibinfo {pages} {149--187} (\bibinfo {year}
		{1996})}\BibitemShut {NoStop}%
	\bibitem [{\citenamefont {Andersen}(2012)}]{andersen2012support}%
	\BibitemOpen
	\bibfield  {author} {\bibinfo {author} {\bibfnamefont {R.}~\bibnamefont
			{Andersen}},\ }\bibfield  {title} {\enquote {\bibinfo {title} {Support for
				democracy in cross-national perspective: The detrimental effect of economic
				inequality},}\ }\href@noop {} {\bibfield  {journal} {\bibinfo  {journal}
			{Research in Social Stratification and Mobility}\ }\textbf {\bibinfo {volume}
			{30}},\ \bibinfo {pages} {389--402} (\bibinfo {year} {2012})}\BibitemShut
	{NoStop}%
	\bibitem [{\citenamefont {Cederman}, \citenamefont {Gleditsch},\ and\
		\citenamefont {Buhaug}(2013)}]{cederman2013inequality}%
	\BibitemOpen
	\bibfield  {author} {\bibinfo {author} {\bibfnamefont {L.-E.}\ \bibnamefont
			{Cederman}}, \bibinfo {author} {\bibfnamefont {K.~S.}\ \bibnamefont
			{Gleditsch}}, \ and\ \bibinfo {author} {\bibfnamefont {H.}~\bibnamefont
			{Buhaug}},\ }\href@noop {} {\emph {\bibinfo {title} {Inequality, grievances,
				and civil war}}}\ (\bibinfo  {publisher} {Cambridge University Press},\
	\bibinfo {year} {2013})\BibitemShut {NoStop}%
	\bibitem [{\citenamefont {Inglehart}\ and\ \citenamefont
		{Norris}(2016)}]{inglehart2016trump}%
	\BibitemOpen
	\bibfield  {author} {\bibinfo {author} {\bibfnamefont {R.~F.}\ \bibnamefont
			{Inglehart}}\ and\ \bibinfo {author} {\bibfnamefont {P.}~\bibnamefont
			{Norris}},\ }\bibfield  {title} {\enquote {\bibinfo {title} {Trump, brexit,
				and the rise of populism: Economic have-nots and cultural backlash},}\
	}\href@noop {} {\bibfield  {journal} {\bibinfo  {journal} {HKS Working paper
				no. RWP16-026}\ } (\bibinfo {year} {2016})}\BibitemShut {NoStop}%
	\bibitem [{\citenamefont {Meleady}, \citenamefont {Seger},\ and\ \citenamefont
		{Vermue}(2017)}]{meleady2017examining}%
	\BibitemOpen
	\bibfield  {author} {\bibinfo {author} {\bibfnamefont {R.}~\bibnamefont
			{Meleady}}, \bibinfo {author} {\bibfnamefont {C.~R.}\ \bibnamefont {Seger}},
		\ and\ \bibinfo {author} {\bibfnamefont {M.}~\bibnamefont {Vermue}},\
	}\bibfield  {title} {\enquote {\bibinfo {title} {Examining the role of
				positive and negative intergroup contact and anti-immigrant prejudice in
				brexit},}\ }\href@noop {} {\bibfield  {journal} {\bibinfo  {journal} {British
				Journal of Social Psychology}\ }\textbf {\bibinfo {volume} {56}},\ \bibinfo
		{pages} {799--808} (\bibinfo {year} {2017})}\BibitemShut {NoStop}%
	\bibitem [{\citenamefont {Jay}\ \emph {et~al.}(2019)\citenamefont {Jay},
		\citenamefont {Batruch}, \citenamefont {Jetten}, \citenamefont {McGarty},\
		and\ \citenamefont {Muldoon}}]{jay2019economic}%
	\BibitemOpen
	\bibfield  {author} {\bibinfo {author} {\bibfnamefont {S.}~\bibnamefont
			{Jay}}, \bibinfo {author} {\bibfnamefont {A.}~\bibnamefont {Batruch}},
		\bibinfo {author} {\bibfnamefont {J.}~\bibnamefont {Jetten}}, \bibinfo
		{author} {\bibfnamefont {C.}~\bibnamefont {McGarty}}, \ and\ \bibinfo
		{author} {\bibfnamefont {O.~T.}\ \bibnamefont {Muldoon}},\ }\bibfield
	{title} {\enquote {\bibinfo {title} {Economic inequality and the rise of
				far-right populism: A social psychological analysis},}\ }\href@noop {}
	{\bibfield  {journal} {\bibinfo  {journal} {Journal of Community \& Applied
				Social Psychology}\ }\textbf {\bibinfo {volume} {29}},\ \bibinfo {pages}
		{418--428} (\bibinfo {year} {2019})}\BibitemShut {NoStop}%
	\bibitem [{\citenamefont {Sniderman}\ and\ \citenamefont
		{Hagendoorn}(2009)}]{sniderman2009ways}%
	\BibitemOpen
	\bibfield  {author} {\bibinfo {author} {\bibfnamefont {P.~M.}\ \bibnamefont
			{Sniderman}}\ and\ \bibinfo {author} {\bibfnamefont {L.}~\bibnamefont
			{Hagendoorn}},\ }\bibfield  {title} {\enquote {\bibinfo {title} {When ways of
				life collide},}\ }in\ \href@noop {} {\emph {\bibinfo {booktitle} {When Ways
				of Life Collide}}}\ (\bibinfo  {publisher} {Princeton University Press},\
	\bibinfo {year} {2009})\BibitemShut {NoStop}%
	\bibitem [{\citenamefont {Garc{\'\i}a-Castro}, \citenamefont
		{Rodr{\'\i}guez-Bail{\'o}n},\ and\ \citenamefont
		{Willis}(2020)}]{garcia2020perceiving}%
	\BibitemOpen
	\bibfield  {author} {\bibinfo {author} {\bibfnamefont {J.~D.}\ \bibnamefont
			{Garc{\'\i}a-Castro}}, \bibinfo {author} {\bibfnamefont {R.}~\bibnamefont
			{Rodr{\'\i}guez-Bail{\'o}n}}, \ and\ \bibinfo {author} {\bibfnamefont
			{G.~B.}\ \bibnamefont {Willis}},\ }\bibfield  {title} {\enquote {\bibinfo
			{title} {Perceiving economic inequality in everyday life decreases tolerance
				to inequality},}\ }\href@noop {} {\bibfield  {journal} {\bibinfo  {journal}
			{Journal of Experimental Social Psychology}\ }\textbf {\bibinfo {volume}
			{90}},\ \bibinfo {pages} {104019} (\bibinfo {year} {2020})}\BibitemShut
	{NoStop}%
	\bibitem [{\citenamefont {Chisadza}, \citenamefont {Nicholls},\ and\
		\citenamefont {Yitbarek}(2021)}]{chisadza2021group}%
	\BibitemOpen
	\bibfield  {author} {\bibinfo {author} {\bibfnamefont {C.}~\bibnamefont
			{Chisadza}}, \bibinfo {author} {\bibfnamefont {N.}~\bibnamefont {Nicholls}},
		\ and\ \bibinfo {author} {\bibfnamefont {E.}~\bibnamefont {Yitbarek}},\
	}\bibfield  {title} {\enquote {\bibinfo {title} {Group identity in fairness
				decisions: Discrimination or inequality aversion?}}\ }\href@noop {}
	{\bibfield  {journal} {\bibinfo  {journal} {Journal of Behavioral and
				Experimental Economics}\ }\textbf {\bibinfo {volume} {93}},\ \bibinfo {pages}
		{101722} (\bibinfo {year} {2021})}\BibitemShut {NoStop}%
	\bibitem [{\citenamefont {Kirkland}\ \emph {et~al.}(2021)\citenamefont
		{Kirkland}, \citenamefont {Jetten}, \citenamefont {Wilks},\ and\
		\citenamefont {Nielsen}}]{kirkland2021economic}%
	\BibitemOpen
	\bibfield  {author} {\bibinfo {author} {\bibfnamefont {K.}~\bibnamefont
			{Kirkland}}, \bibinfo {author} {\bibfnamefont {J.}~\bibnamefont {Jetten}},
		\bibinfo {author} {\bibfnamefont {M.}~\bibnamefont {Wilks}}, \ and\ \bibinfo
		{author} {\bibfnamefont {M.}~\bibnamefont {Nielsen}},\ }\bibfield  {title}
	{\enquote {\bibinfo {title} {How economic inequality affects prosocial
				behavior in children across development},}\ }\href@noop {} {\bibfield
		{journal} {\bibinfo  {journal} {Journal of Experimental Child Psychology}\
		}\textbf {\bibinfo {volume} {210}},\ \bibinfo {pages} {105202} (\bibinfo
		{year} {2021})}\BibitemShut {NoStop}%
	\bibitem [{\citenamefont {Koster}\ \emph {et~al.}(2022)\citenamefont {Koster},
		\citenamefont {Balaguer}, \citenamefont {Tacchetti}, \citenamefont
		{Weinstein}, \citenamefont {Zhu}, \citenamefont {Hauser}, \citenamefont
		{Williams}, \citenamefont {Campbell-Gillingham}, \citenamefont {Thacker},
		\citenamefont {Botvinick} \emph {et~al.}}]{koster2022human}%
	\BibitemOpen
	\bibfield  {author} {\bibinfo {author} {\bibfnamefont {R.}~\bibnamefont
			{Koster}}, \bibinfo {author} {\bibfnamefont {J.}~\bibnamefont {Balaguer}},
		\bibinfo {author} {\bibfnamefont {A.}~\bibnamefont {Tacchetti}}, \bibinfo
		{author} {\bibfnamefont {A.}~\bibnamefont {Weinstein}}, \bibinfo {author}
		{\bibfnamefont {T.}~\bibnamefont {Zhu}}, \bibinfo {author} {\bibfnamefont
			{O.}~\bibnamefont {Hauser}}, \bibinfo {author} {\bibfnamefont
			{D.}~\bibnamefont {Williams}}, \bibinfo {author} {\bibfnamefont
			{L.}~\bibnamefont {Campbell-Gillingham}}, \bibinfo {author} {\bibfnamefont
			{P.}~\bibnamefont {Thacker}}, \bibinfo {author} {\bibfnamefont
			{M.}~\bibnamefont {Botvinick}},  \emph {et~al.},\ }\bibfield  {title}
	{\enquote {\bibinfo {title} {Human-centered mechanism design with democratic
				{AI}},}\ }\href@noop {} {\bibfield  {journal} {\bibinfo  {journal} {Nature
				Human Behaviour}\ } (\bibinfo {year} {2022})}\BibitemShut {NoStop}%
\end{thebibliography}

%

\end{document}